\documentclass{aa}
\thesaurus{08(08.01.2;08.03.3;08.06.3;08.16.2;03.20.8)}
\usepackage{txfonts}
\usepackage{graphicx,natbib}
\bibpunct{(}{)}{;}{a}{}{,}
\begin{document}
\title{On the Ages of Exoplanet Host Stars
\thanks{Based on observations collected at
the Complejo Astron\'omico El Leoncito (CASLEO) operated under agreement
between the Consejo Nacional
de Investigaciones Cient\'\i ficas y T\'ecnicas de la Rep\'ublica
Argentina and the National
Universities of La Plata, C\'ordoba and San Juan.}}

\author{Carlos Saffe\fnmsep\thanks{On a fellowship from CONICET,
Argentina.}, Mercedes G\'omez and Carolina Chavero\fnmsep\thanks{Also at
the Facultad de Matem\'atica, Astronom\'\i a y F\'\i sica de la Universidad
Nacional de C\'ordoba, Argentina.}\fnmsep\thanks{Present address:
Observat\'orio Nacional, Rua General Jos\'e Cristino 77, S\~ao Cristov\~ao,
20921-400 Rio de Janeiro, Brasil.}}

\institute{Observatorio Astron\'omico de C\'ordoba, Laprida 854, 5000
C\'ordoba, Argentina \\ email: saffe@oac.uncor.edu, mercedes@oac.uncor.edu,
carolina@oac.uncor.edu}

\offprints{C. Saffe}

\authorrunning{Saffe et al.}

\titlerunning{Exoplanet Host Stars: Ages}

\date{Received Month XX, 200X; accepted Month XX, 200X}

\maketitle

\abstract{
We obtained spectra, covering the CaII H and K region, for 49 exoplanet host
(EH) stars, observable from the southern hemisphere. We measured the
chromospheric activity index, R$'$${_{\rm HK}}$. We compiled previously
published values of this index for the observed objects as well as the
remaining EH stars in an effort to better smooth temporal variations
and derive a more representative value of the
average chromospheric activity for each object. We used
the average index to obtain ages for the group of EH stars.
In addition we applied other methods, such as:
Isochrone, lithium abundance, metallicity and transverse velocity
dispersions, to compare with the chromospheric results.
The kinematic method is a less reliable age estimator because EH stars
lie red-ward  of Parenago's discontinuity in the transverse velocity
dispersion vs dereddened B$-$V diagram.
The chromospheric and isochrone techniques give median ages
of \hbox{5.2 and 7.4 Gyr,} respectively, with a dispersion of $\sim$ 4 Gyr.
The median age of F and G EH stars derived by the
isochrone technique is $\sim$ 1--2 Gyr older than that of identical spectral type
nearby stars not known to be associated with planets.  However,
the dispersion in both cases is large, about $\sim$ 2--4 Gyr.
We searched for correlations between the chromospheric and isochrone ages and
L$_{\rm IR}$/L$_{\rm *}$ (the excess over the stellar luminosity) and
the metallicity of the EH stars. No clear tendency is found in the first
case, whereas the metallicy dispersion seems to slightly increase with age.


}

\authorrunning{Saffe et al.}

%
%

\section{Introduction}

During the last decade, the detection of more than one hundred nearby solar-type
stars associated with likely single or multiple planetary mass companions
\citep{mq95,butler99} has given rise to new interest in the study of
these relatively bright stellar objects \citep[see, for example,][]{reid}.

At the present time, most of the known EH stars have been
detected by means of the Doppler technique and
are, in general, among the less chromospherically active and
slow-rotation solar-type stars. The reason for this
selection effect is that
chromospherically active stars have stellar
surface features, such as convective inhomogeneities
or magnetic spots that may induce intrinsic stellar
radial-velocity ''jitter'' indistinguishable from
the orbital motion of the star around the center of mass of the star and
planet system \citep{sado97,saar98}.  These effects may inhibit or even
provide false detections \citep{wal92,san00b,que01,pau02,pau04}.

It is well established that the Exoplanet Host (EH) sample, on
average, is metal-rich compared to solar neighborhood field stars not known to have
planets, detectable by means of high precision radial velocity
measurements \citep[][]{gon1,la97,gon98,gon3,san1,san2,santos}.

\cite{sush} analyzed nine F-type stars associated with exoplanets and
determined their ages by using different estimators, such as: metallicity,
Hipparcos variability, brightness anomaly and location in the color-magnitude
diagram in relation to field F stars and the Hyades. They concluded that
the 9 analyzed stars have ages similar to Hyades ($\sim$ 0.7 Gyr) and thus
are significantly younger than F field stars. The age may thus
be a parameter that can help in selecting candidate EH stars.
Moreover, the age is a fundamental parameter that it is worth while
exploring.

In this contribution we present spectra for 49 EH stars observed
from the southern hemisphere and apply the
stellar chromospheric activity to obtain their ages. From the
literature we derive the chromospheric index, R$'$${_{\rm HK}}$, for
the remaining stars with no spectra reported in this contribution.
We compare the ''chromospheric'' age determinations with those calculated using
other methods, such as: Isochrones, lithium and metallicity abundances, and
space velocity dispersions.  We confront the EH stars ages with those of nearby
stars of similar characteristics not known to be associated with planets.
We search for correlations between the age and physical parameters of
the EH stars such as, the L$_{\rm IR}$/L$_{\rm *}$
(the excess over the stellar luminosity) and the metallicity.

In Section 2 we present our observations and in Section 3 we apply the CaII H, K core
emissions to measure the chromospheric activity and to derive ages.
The other age estimators are described and discussed in Sections 4.
We compare the ages of the EH stars with those of solar neighborhood stars
of similar spectral types not associated with planets in Section 5.
Finally, we search for correlations of physical properties of the stars
with age in Section 6.  We conclude with a brief summary in Section 7.

\section{Observations and data reduction}

We observed 49 southern hemisphere EH stars
from the California and Carnegie Planet Search\footnote{http://exoplanets.org}
and the Geneva Observatory Planet Search
\footnote{http://obswww.unige.ch/exoplanets} lists.
These compilations basically include 138 EH stars up to 06/25/2005,
including 157 exoplanets and 14 multiple systems.
131 EH stars have been
detected by Doppler spectroscopy and only 7 by photometry. The likely planetary
companions have masses such that M sin {\it i} $<$ 17 M$_{\rm JUP}$.  The 49
stars we observed have distances between \hbox{10 and 94 pc} and spectral types
F, G, and K (6, 34, and 9 objects, respectively), as specified in the Hipparcos
database.

We carried out the observations on September 20--22 2003 and March 28--31 2004,
at the Complejo Astronomico El Leoncito (CASLEO, San Juan - Argentina) with the
REOSC spectrograph attached to the Jorge Sahade 2.15-m telescope. The REOSC has
a TEK 1024 $\times$ 1024 back illuminated detector, with a pixel size of
24 $\times$ 24~$\mu$m, and it was employed in single dispersion mode.
We used a 1200 l/mm grating (0.75 {\AA}/pix) centered at 3950 {\AA}
to cover the spectral range 3500--4200 {\AA}, including the
CaII H and K lines, at 3968 and 3933 {\AA}, respectively. We selected
a 250~ $\mu$m ($\sim$ 1$''$) wide slit.  Several chromospheric ''standards''
were observed during both observing runs.  The integration times varied between
1 and 10 minutes, depending on the sources brightness. A pair of CuNeAr lamp
spectra was taken for each object.  To reduce the spectra and
measure CaII H and K lines fluxes we used IRAF\footnote{IRAF is distributed
by the National Optical Astronomy Observatory, which is operated by
the Association of Universities for Research in Astronomy, Inc.
under contract to the National Science Foundation.}.

The spectra were extracted using
the NOAO task {\it apall} with an aperture of 5 pixel radius.
A sky subtraction was carried out by fitting a polynomial to the regions on
either side of the aperture. A non linear low order fit to the lines in the
CuNeAr lamp was used to wavelength calibrate the spectra. Typical RMS for
the wavelength solution is 0.22. The {\it sbands} task was used
to measure the fluxes in the CaII lines cores.

\section{Age derivation from the chromospheric activity}

The stellar chromospheric emission (CE) as measured by the core emission in
the CaII H and K absorption lines, is related to both the spectral type \citep{st1,bal1}
and the rotational velocity of the central star
\citep[see, for example,][]{wil,sku,barry,eggen,soder}.
Late spectral type main-sequence stars have larger chromospheric activity
than early type objects \citep{st1}.  As the object ages, it slows down its rotation
and diminishes the level of CE \citep[][]{wil,noyes,rm1}. In this sense,
the CE provides an indication of the stellar age for a given spectral type.

The chromospheric activity is quantified by the S and R$'$$_{\rm HK}$ indexes
\citep[e.g. ][]{vp2,bal1,vp,bal2,soder,hen}.
The S index is defined by the sum of fluxes within two 1-{\AA}-width bands
centered on the CaII H ($\lambda$3968 {\AA}) and K
($\lambda$3933 {\AA}) lines.  Then the combined flux is normalized to the
pseudo-continuum level as measured by two equidistant windows of 20-{\AA}-width
each, on either side of the CaII lines. According to this definition
it is not necessary to flux calibrate the spectra as the index definition involved
relative measurements.

The R$'$$_{\rm HK}$ index introduces two modifications to the S index: 1)
a B$-$V color correction,
\vskip 0.1in

\begin{equation}
{\rm R_{HK} = C(B-V)~S,}
\label{one}
\end{equation}
\vskip 0.1in

\noindent
and 2) the substraction of the photospheric contribution, R$_{\rm PHOT}$:
\vskip 0.1in

\begin{equation}
{\rm R'_{HK} =  R_{HK} - R_{PHOT}.}
\label{dos}
\end{equation}
\vskip 0.1in

\noindent
We refer to the paper of \cite{noyes} for a detailed description of
the derivation of both C(B$-$V) and R$_{\rm PHOT}$.

To determine S and R$'$${_{\rm HK}}$ for the observed stars we basically
adopted the procedure of \cite{hen}. We define the S$_{\rm CASLEO}$ index
analogous to Equation \ref{one} and transform this index to the Mount Wilson four
spectrophotometric bands \citep{vp2} by means of the standard stars measurements.
Specifically, in the determination of S$_{\rm CASLEO}$ we used two 3-{\AA}-width
bands centered on the CaII lines and two 20-{\AA}-width pseudo-continuum windows
located on either side of the H and K lines. In Table \ref{stand} we list the
standard stars observed, selected from among those with more that 100 observations
at Mount Wilson. We include the S and R$'$${_{\rm HK}}$ indexes corresponding to
the Mount Wilson and the CASLEO measurements.

To estimate the errors in our determinations of S$_{\rm CASLEO}$ for the
observed objects we displaced
the on-line windows half a pixel on either direction, re-calculated the index
in each case and then compared with the original measurements.  In this manner
we estimate an
error of $\sim$ 0.005 in S$_{\rm CASLEO}$ due to misplacements of the on-line
windows. Translated to age, an error of $\sim$ 0.005 corresponds to $\sim$ 0.4 Gyr
for a 5 Gyr old star. In this estimation we used the \cite{don}'s calibration.

\begin{table*}
\center
\caption{Chromospheric standard stars measured at the CASLEO.}
\begin{tabular}{lcccc}
\hline
Star         &S$_{\rm MW}$& Log R$'$$_{\rm MW}$ &  S$_{\rm CASLEO}$ &  Log R$'$$_{\rm CASLEO}$ \\
\hline
             &        & September 2003 &               &         \\
\hline
HD 3443 AB   & 0.1823 & $-$4.907         &  0.3706       &  $-$5.077  \\
HD 3795      & 0.1557 & $-$5.038         &  0.3443       &  $-$5.146  \\
HD 9562      & 0.1365 & $-$5.174         &  0.3526       &  $-$5.124  \\
HD 10700     & 0.1712 & $-$4.959         &  0.3739       &  $-$5.066  \\
HD 11131     & 0.3355 & $-$4.428         &  0.4582       &  $-$4.580  \\
HD 16673     & 0.2151 & $-$4.662         &  0.4307       &  $-$4.671  \\
HD 17925     & 0.6478 & $-$4.314         &  0.5870       &  $-$4.263  \\
HD 22049     & 0.4919 & $-$4.458         &  0.5129       &  $-$4.536  \\
HD 30495     & 0.2973 & $-$4.510         &  0.4431       &  $-$4.648  \\
HD 38393     & 0.1514 & $-$4.941         &  0.3842       &  $-$4.946  \\
HD 152391    & 0.3867 & $-$4.461         &  0.5053       &  $-$4.432  \\
HD 158614 AB & 0.1581 & $-$5.028         &  0.3708       &  $-$5.076  \\
\hline
             &        & March 2004       &               &          \\
\hline
HD 23249     & 0.1374 & $-$5.184        &  0.2754       &  $-$5.037  \\
HD 30495     & 0.2973 & $-$4.510        &  0.4007       &  $-$4.414  \\
HD 38392     & 0.5314 & $-$4.497        &  0.4811       &  $-$4.389  \\
HD 38393     & 0.1514 & $-$4.941        &  0.3385       &  $-$5.004  \\
HD 45067     & 0.1409 & $-$5.092        &  0.3178       &  $-$5.094  \\
HD 76151     & 0.2422 & $-$4.670        &  0.3712       &  $-$4.700  \\
HD 81809 AB  & 0.1720 & $-$4.923        &  0.3135       &  $-$5.089  \\
HD 158614    & 0.1581 & $-$5.028        &  0.3138       &  $-$4.944  \\
\hline
\end{tabular}
\label{stand}
\end{table*}

\begin{figure*}
\center
\includegraphics[width=60mm]{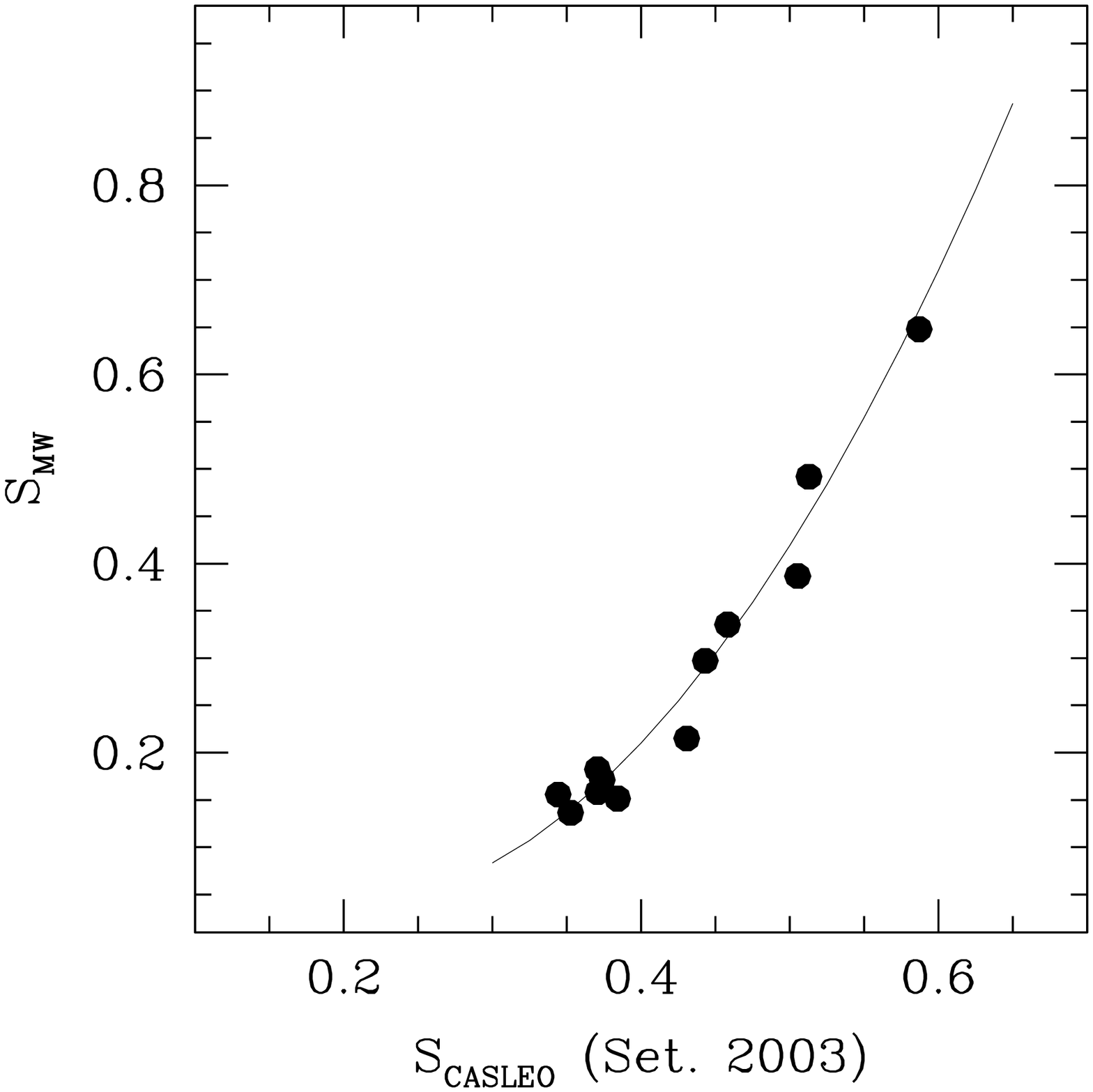}
\includegraphics[width=60mm]{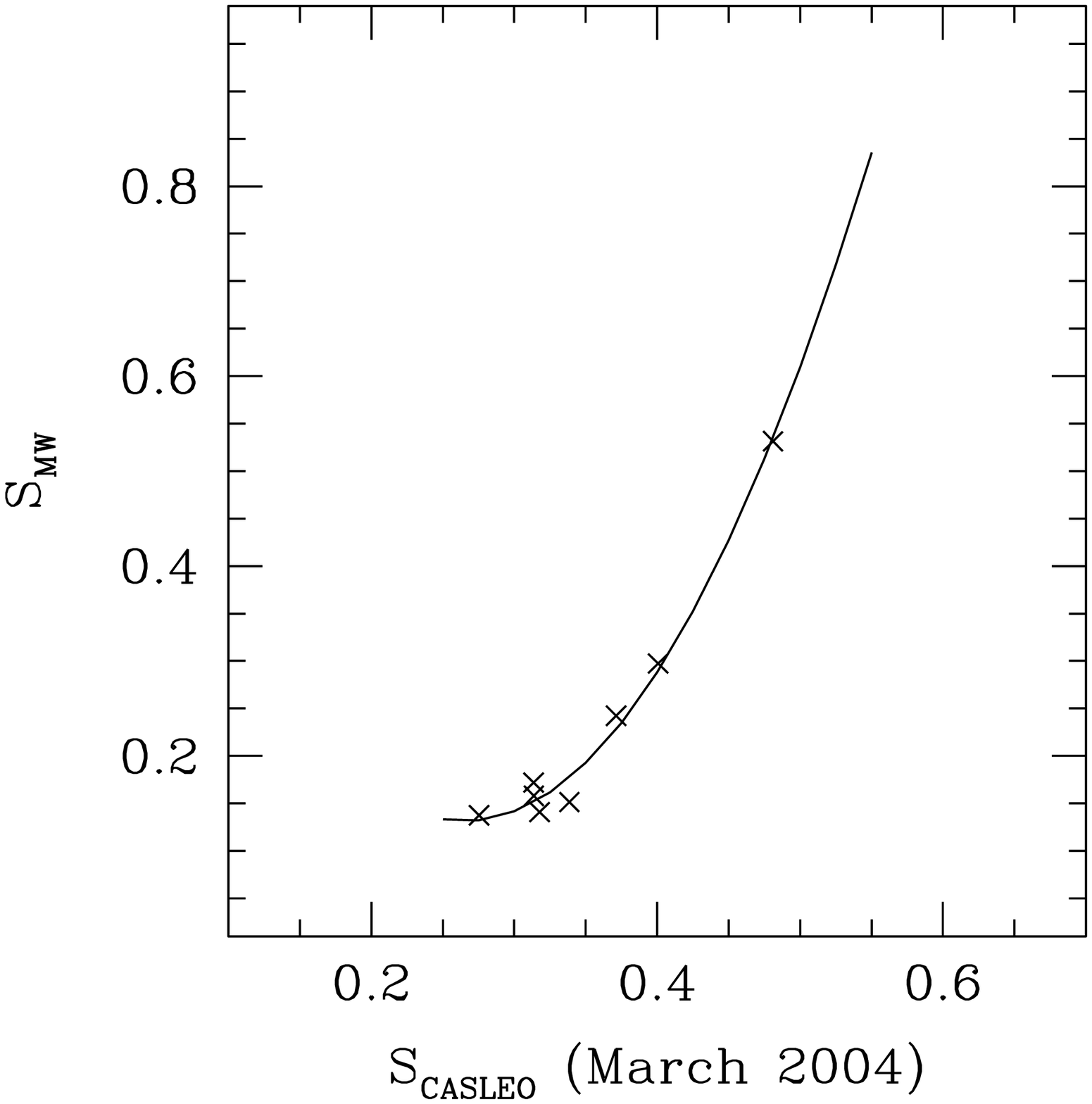}
\includegraphics[width=60mm]{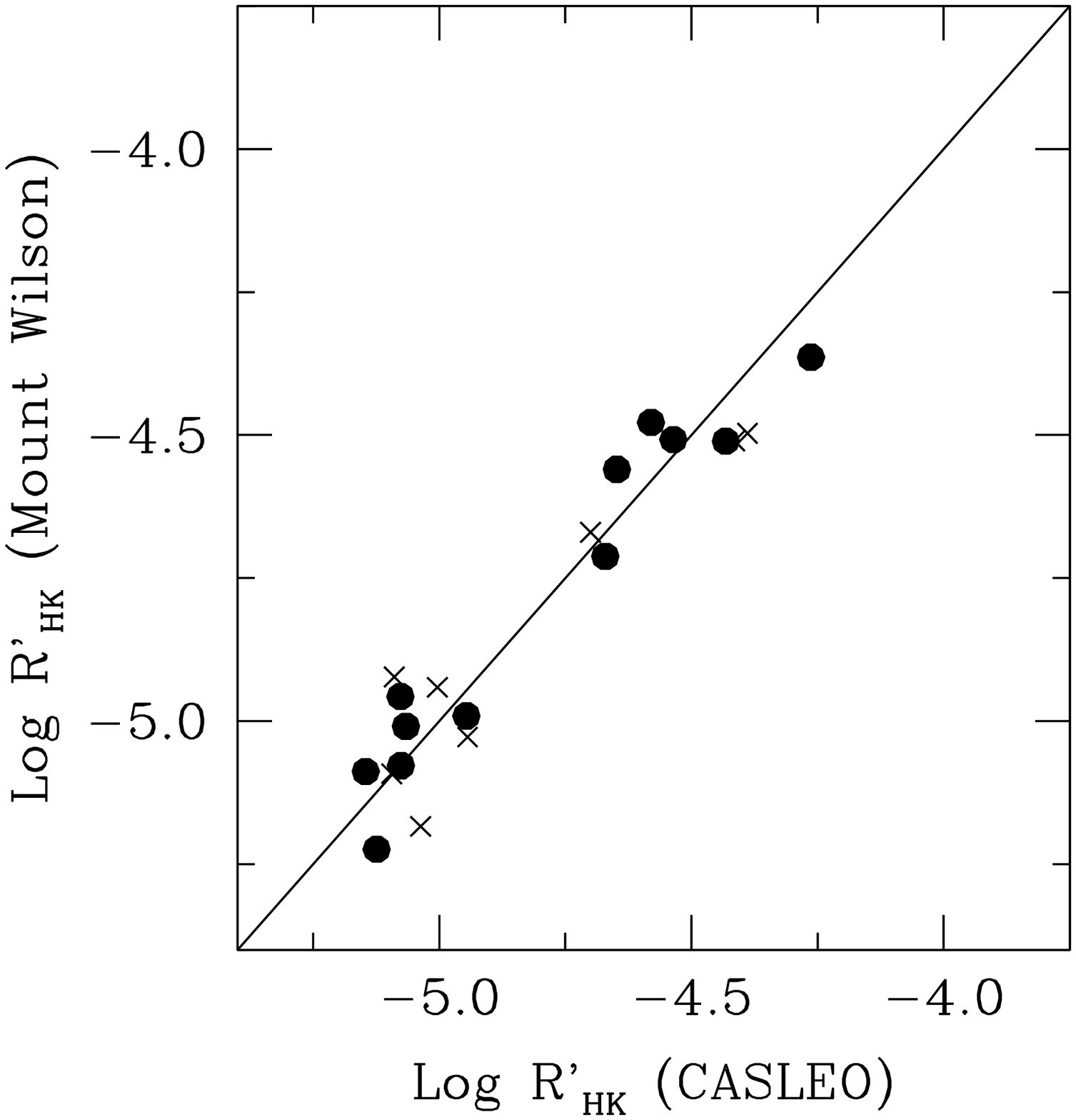}
\caption{Upper panel: S$_{\rm CASLEO}$ vs S$_{\rm MW}$ (MW: Mount Wilson)
for the ''standard'' stars in Table \ref{stand} corresponding to the September
2003 and March 2004 observing runs. Lower panel: Log R$'$$_{\rm HK}$ values for both
observing sites (i.e., CASLEO and Mount Wilson). The open circles indicate 2003 data
and the crosses 2004 observations.}
\label{std1}
\end{figure*}

Figure \ref{std1}, upper panel, shows the S$_{\rm CASLEO}$ vs S$_{\rm MW}$ indexes plot
corresponding to each observing run.  We used a second order fit to reproduce
the data point (the continuous line) and derive the following relations:
\vskip 0.1in

\begin{equation}
{\rm S_{MW} = 4.1109~~ S_{CASLEO}^{2} - 1.6104~~ S_{CASLEO} + 0.1966}
\label{tres}
\end{equation}
\vskip 0.1in

\begin{equation}
{\rm S_{MW} = ~8.7210~~ S_{CASLEO}^{2} - 4.6370~~ S_{CASLEO} + 0.7476},
\label{cuatro}
\end{equation}
\vskip 0.1in

\noindent
for the September 2003 and March 2004 observing runs, respectively.
These relations are strictly valid for 0.27 $<$ S$_{\rm CASLEO}$ $<$ 0.59
(0.14 $<$ S$_{\rm MW}$ $<$ 0.65) September 2003 and \hbox{0.27 $<$ S$_{\rm CASLEO}$ $<$ 0.48}
(0.14 $<$ S$_{\rm MW}$ $<$ 0.53) March 2004.  HD 162020, one of the most
chromospheric active EH stars (S$_{\rm CASLEO}$ $=$ 1.11), is
the only object in our sample outside the ranges of Equations
\ref{tres} and \ref{cuatro}. In this case we extrapolated these relations
to include this object in our analysis.

Figure \ref{std1}, lower panel, compares
the Log R$'$$_{\rm HK}$ values corresponding to CASLEO and Mount Wilson for
the standard stars measurements (see also Table \ref{stand}).  We derived an
uncertainty of \hbox{$\sim$ 0.05 dex} for the CASLEO calibration with respect to
the Mount Wilson relation. This value mainly reflects the fact that the CE of
the stars varies over time. Systematic errors in the CASLEO calibration with respect
to the Mount Wilson standard are likely to be much smaller than this amount. An
uncertainty of $\sim$ 0.05 dex, similar to those derived by \cite{hen} or \cite{R33},
corresponds to an age difference of $\sim$ 1.5 Gyr for a 5 Gyr old star,
using the \cite{don}'s calibration.

The CE varies with time, having short and long periodic
and non-periodic variations \citep[][]{noyes, bal1, bal2}.
The use of instantaneous values of the indexes S and Log R$'$$_{\rm HK}$ (i.e.,
corresponding to a given epoch of observation) can induce to erroneous age
estimations. For example, in the case of the Sun the R$'$$_{\rm HK}$ varied
from $-$4.75 to $-$5.10 during the ''Maunder Minimum'' ($\sim$ 1650, $\sim$ 1890),
corresponding to ages of 8.0 and 2.2 Gyr, respectively. Although this
represents an extreme variation (at present, only certainly detected for the Sun),
it cautions on the applicability of individual values of the CE index. For
this reason it is more appropriate to use the temporal average,
$<$R$'$$_{\rm HK}$$>$, to estimate the age.

We have extensively searched the literature for previous determinations
of the index R$'$$_{\rm HK}$ for all the stars observed at the CASLEO and
the rest of the sample of EH stars.  Table \ref{cromoref} shows a one-sample
page of our compilation as well as the data reported in this contribution.
The complete table is available in electronic format.  As the data are published
in different manners, for example individual or average observations for each
observing run, we have also indicated the type of data
used in the final average in Table \ref{cromoref}.

\begin{table*}
\center
\caption{Chromospheric index compilation for the EH stars: A sample page}
\begin{tabular}{lllll}
\hline
       Name        &    S    & Log R$'$$_{\rm HK}$ &   Data Type                  & Ref \\
\hline
HD 10697           & 0.1279  &       & 1 obs in 1 day (01/09/78 to 01/09/78)     & R19 \\
                   & 0.1423  &       & 2 obs in 10 days (15/11/79 to 24/11/79)    & R19 \\
                   &         & $-$5.02 & 35 obs in 3 years                          & R34 \\
                   & 0.149   & $-$5.08 & 57 obs in 25 month bins                    & R52 \\
\hline
HD 12661           &         & $-$5.00 & 1 obs 1998$-$99                              & R33 \\
                   & 0.14    & $-$5.12 & ~30 obs in 2 years                         & R38 \\
                   & 0.150   & $-$5.08 & 52 obs in 16 month bins                    & R52 \\
\hline
HD 16141           & 0.145   &       & 1979 individual                            & R05 \\
                   & 0.145   &       & R1; R6; R5; R10; R3                        & R17 \\
                   & 0.1452  &       & 1 obs in 1 day (15/11/79 to 15/11/79)     & R19 \\
                   &         & $-$5.05 & 46 obs in 4 years                          & R37 \\
                   & 0.145   & $-$5.11 & 70 obs in 23 month bins                    & R52 \\
                   & 0.085   &       & individual                                 & CASLEO \\
\hline
HD 17051           & 0.225   & $-$4.65 & 1992 individual                            & R28 \\
                   &         & $-$4.65 &                                            & R66 \\
                   & 0.2074  &       & individual                                 & CASLEO \\
\hline
HD 19994           & 0.173   & $-$4.88 & 12 obs in 4 month bins                     & R52 \\
                   &         & $-$4.77 & ~45 obs in 6 years                         & R53 \\
                   & 0.1018  &       & individual                                 & CASLEO\\
\hline
HD 20367           & 0.282   & $-$4.50 & 2 obs in 1 month bins                      & R52 \\
\hline
HD 23079           & 0.164   & $-$4.94 & 1992 individual                            & R28 \\
                   &         & $-$4.96 & individual                                 & R41 \\
                   & 0.1196  &       & individual                                 & CASLEO\\
\hline
HD 23596           & 0.150   & $-$5.06 & 1 obs in 1 month bins                      & R52 \\
\hline
HD 27442           & 0.062   &       & individual                                 & CASLEO\\
\hline
HD 28185           &         & $-$5.00 & 1 obs 1998$-$99                              & R33 \\
                   & 0.143   &       & individual                                 & CASLEO\\
\hline
HD 30177           &         & $-$5.08 & individual                                 & R41 \\
                   & 0.1205  &       & individual                                 & CASLEO\\
\hline
HD 33636           &         & $-$4.81 & 21 obs in 3 years                          & R45 \\
                   & 0.180   & $-$4.85 & 25 obs in 13 month bins                    & R52 \\
                   & 0.135   &       & individual                                 & CASLEO\\
\hline
\end{tabular}
\label{cromoref}
\end{table*}

\setcounter{table}{1}
\begin{table*}
\center
\caption{Continued. References}
\begin{tabular}{lll}
\hline
CASLEO: this paper           &  R23: \cite{R23}           &R46: \cite{R46}              \\
R01: \cite{vp}               &  R24: \cite{R24}           &R47: \cite{marcy3}           \\
R02: \cite{R02}              &  R25: \cite{R25}           &R48: \cite{R48}              \\
R03: \cite{R03}              &  R26: \cite{R26}           &R49: \cite{butler3}          \\
R04: \cite{R04}              &  R27: \cite{R27}           &R50: \cite{R50}              \\
R05: \cite{R05}              &  R28: \cite{hen}           &R51: \cite{R51}              \\
R06: \cite{R06}              &  R29: \cite{R29}           &R52: \cite{wright2}          \\
R07: \cite{R07}              &  R30: \cite{R30}           &R53: \cite{R53}              \\
R08: \cite{R08}              &  R31: \cite{R31}           &R54: \cite{R54}              \\
R09: \cite{R09}              &  R32: \cite{fischer1}      &R55: \cite{R55}              \\
R10: \cite{noyes}            &  R33: \cite{R33}           &R56: \cite{R56}              \\
R11: \cite{R11}              &  R34: \cite{R34}           &R57: \cite{R57}              \\
R12: \cite{R12}              &  R35: \cite{R35}           &R58: \cite{R58}              \\
R13: \cite{lacha}            &  R36: \cite{R36}           &R59: \cite{R59}              \\
R14: \cite{R14}              &  R37: \cite{R37}           &R60: \cite{R60}              \\
R15: \cite{R15}              &  R38: \cite{fischer2}      &R61: \cite{san00b}          \\
R16: \cite{R16}              &  R39: \cite{R39}           &R62: \cite{R62}              \\
R17: \cite{R17}              &  R40: \cite{R40}           &R63: \cite{R63}              \\
R18: \cite{R18}              &  R41: \cite{R41}           &R64: \cite{R64}              \\
R19: \cite{R19}              &  R42: \cite{R42}           &R65: \cite{R65}              \\
R20: \cite{soder}            &  R43: \cite{R43}           &R66: \cite{rm1}              \\
R21: \cite{R21}              &  R44: \cite{R44}           &                             \\
R22: \cite{bal1}             &  R45: \cite{R45}           &                             \\
\hline
\end{tabular}
\end{table*}

In Table \ref{cromoages} we list the Log R$'$$_{\rm HK}$ values derived from the
CASLEO data only,
the average compiled from the literature, $<$Log R$'$$_{\rm HK_{~ without~ CASLEO}}$$>$,
and the final average including our CASLEO measurements,
$<$Log R$'$$_{\rm HK_{~ with~ CASLEO}}$$>$.  The average difference between the indexes with
and without the data reported in this contribution is $\sim$ 0.107.  Three stars in the
sample, \hbox{HD 19994}, \hbox{HD 169830} and \hbox{HD 216437}, show average differences
significantly larger, 0.569, 0.381 and 0.222, respectively. In Table \ref{cromoagesnotcasleo}
we list the $<$Log R$'$$_{\rm HK}$$>$ obtained from the literature for the objects not observed
at the CASLEO.

To derive the ''chromospheric age'' for the EH objects we applied the
calibrations of \cite{don}, hereafter D93, and \cite{rm1},
hereafter RPM98.  The latter relation includes a correction in the age derivation
due to the stellar metallicity. The derived values are listed
in columns 5 and 6 of Table \ref{cromoages} for the stars observed at the
CASLEO and in columns 3 and 4 of Table \ref{cromoagesnotcasleo} for the objects
with chromospheric activity index, Log R$'$$_{\rm HK}$, compiled from the
literature.

These calibrations are strictly valid for chromospherically quiet
\citep[i.e., \hbox{Log R$'$$_{\rm HK}$ $<$ $-$ 4.75;}][]{vp} late-type F and
G dwarfs. Moreover, \cite{wright1} showed that the canonical
chromospheric activity-age relation breaks down for the less
active stars (i.e., with Log R$'$$_{\rm HK}$  $<$ $-$5.1). In our
case, 66\% of the sample (74 out of 112 stars) in
Tables \ref{cromoages} and \ref{cromoagesnotcasleo} has
$-$5.1 $<$ Log R$'$$_{\rm HK}$ $<$ $-$ 4.75, whereas 38 objects
(i.e., 34\% of the sample) have Log R$'$$_{\rm HK}$ values
outside this range.  The latter objects
may have less reliable chromospheric age determinations.

Adopting the D93 calibration the range of ages corresponding
to the above Log R$'$$_{\rm HK}$ limits, goes from $\sim$ 2.2 to
5.6 Gyr.  However, the D93 calibration has been used
beyond the \cite{wright1}'s limit of \hbox{Log R$'$$_{\rm HK}$ $=$ $-$5.1}
\citep[see, for example,][]{hen97,R31,donahue,henry2000,pepe,wright2}.

\citet{pace} used high-resolution spectra to
derive the chromospheric activity-age relationship
for a set of 5 clusters, spanning a range of age from 0.6 to 4.5 Gyr.
Specifically the group of clusters analyzed by these authors includes
two young objects (Hyades and Praesepe with ages of $\sim$ 0.6 Gyr),
two intermediate age clusters (IC 4651 and NGC 3680 with ages of $\sim$ 1.7 Gyr)
and M 67, a relatively old object (age of $\sim$ 4.5 Gyr). They obtained
spectroscopic data for 21 stars which belong to the young clusters, 7 stars in
the intermediate age objects and 7 stars in M 67.  They found that the
two intermediate age clusters show similar Ca II K activity level to the older
M 67 and the Sun itself. The chromospheric activity-age relationship seems to
decrease very rapidly between 0.6 and about 2 Gyr, after which it enters in a plateau.

This result imposes a serious limitation on the applicability of the chromospheric
technique to derive ages for relatively old stars, with ages $>$ 2 Gyr. In particular
in the case of the EH stars, 85\% of the sample (95 out of 112) has ages older than
the above limit, using D93's calibration (see Tables \ref{cromoages} and
\ref{cromoagesnotcasleo}), and thus the
chromospheric activity as age indicator would have little practical use.

\citet{pace}'s stellar sample is relatively small. In particular their
result is based on 7 high resolution spectra of intermediate age stars. It would
be desirable to extend this analysis to include additional objects per cluster
and a relatively larger number of clusters. This would help to discard any
peculiarity in IC 4651 and NGC 3680 and put \citet{pace}'s results on more
statistically solid grounds. On the other hand, \citet{wright1}'s analysis is based
on roughly 3000 near-by stars, one third of which has high resolution data.
For the time being and in view of \citet{pace}'s result we will indicate how
the 2 Gyr cut-off in the chromospheric activity age relation affects our analysis
for the EH stars.

\begin{table*}
\center
\caption{Chromospheric index, Log R$'$$_{\rm HK}$, and age for the EH stars observed
at the CASLEO}
\begin{tabular}{lccccc}
\hline
Name                  & Log R$'$$_{\rm HK_{~ CASLEO}}$ & $<$Log R$'$$_{\rm HK_{~ without~ CASLEO}}$$>$ &
$<$Log R$'$$_{\rm HK_{~ with~ CASLEO}}$$>$  & D93 Age[Gy]  & RPM98 Age[Gy] \\
\hline
GJ 86           &    $-$4.67&    $-$4.74&    $-$4.72&     2.03&     2.94\\
HD 142          &    $-$5.11&    $-$4.92&    $-$5.02&     5.93&     2.43\\
HD 1237         &    $-$4.31&    $-$4.36&    $-$4.34&     0.15&     0.25\\
HD 2039         &    $-$5.06&    $-$4.91&    $-$4.98&     5.28&     1.20\\
HD 4208         &    $-$4.94&    $-$4.94&    $-$4.94&     4.47&     6.03\\
HD 6434         &    $-$5.23&    $-$4.89&    $-$5.06&     6.85&    18.51\\
HD 17051        &    $-$4.58&    $-$4.65&    $-$4.63&     1.47&     0.43\\
HD 19994        &    $-$5.76&    $-$4.83&    $-$5.14&     8.91&     2.56\\
HD 23079        &    $-$5.23&    $-$4.95&    $-$5.04&     6.53&     5.92\\
HD 27442        &    $-$5.57&         &    $-$5.57&    24.74&     7.15\\
HD 28185        &    $-$4.98&    $-$5.00&    $-$4.99&     5.36&     1.69\\
HD 30177        &    $-$5.15&    $-$5.08&    $-$5.12&     8.30&     1.50\\
HD 33636        &    $-$5.03&    $-$4.83&    $-$4.90&     3.83&     3.24\\
HD 38529        &    $-$5.07&    $-$4.93&    $-$4.97&     5.09&     0.89\\
HD 39091        &    $-$4.82&    $-$4.97&    $-$4.90&     3.83&     1.83\\
HD 52265        &    $-$4.90&    $-$4.97&    $-$4.96&     4.88&     1.65\\
HD 72659        &    $-$4.79&    $-$5.01&    $-$4.94&     4.42&     2.62\\
HD 73526        &    $-$5.00&         &    $-$5.00&     5.59&     1.49\\
HD 75289        &    $-$4.94&    $-$4.98&    $-$4.97&     4.96&     1.29\\
HD 76700        &    $-$4.94&         &    $-$4.94&     4.51&     0.77\\
HD 82943        &    $-$4.77&    $-$4.87&    $-$4.84&     3.08&     0.72\\
HD 83443        &    $-$4.79&    $-$4.85&    $-$4.83&     2.94&     0.63\\
HD 92788        &    $-$4.95&    $-$4.88&    $-$4.89&     3.78&     0.87\\
HD 108147       &    $-$4.64&    $-$4.75&    $-$4.71&     1.98&     0.70\\
HD 114386       &    $-$4.74&         &    $-$4.74&     2.19&     1.91\\
HD 114729       &    $-$4.67&    $-$5.04&    $-$4.95&     4.58&     6.35\\
HD 114783       &    $-$4.70&    $-$4.98&    $-$4.89&     3.70&     1.82\\
HD 121504       &    $-$4.67&    $-$4.65&    $-$4.66&     1.62&     0.66\\
HD 130322       &    $-$4.63&    $-$4.56&    $-$4.58&     1.24&     0.77\\
HD 134987       &    $-$5.13&    $-$5.05&    $-$5.08&     7.32&     1.77\\
HD 141937       &    $-$4.77&    $-$4.80&    $-$4.79&     2.55&     1.25\\
HD 142415       &    $-$4.69&    $-$4.61&    $-$4.63&     1.49&     0.51\\
HD 147513       &    $-$4.40&    $-$4.46&    $-$4.45&     0.65&     0.45\\
HD 160691       &    $-$5.10&    $-$5.02&    $-$5.04&     6.41&     1.45\\
HD 162020       &    $-$4.12&         &    $-$4.12&     0.00&     0.23\\
HD 168443       &    $-$5.00&    $-$5.02&    $-$5.02&     5.90&     3.13\\
HD 168746       &    $-$4.92&    $-$4.87&    $-$4.89&     3.75&     3.18\\
HD 169830       &    $-$5.05&    $-$4.94&    $-$4.97&     4.95&     1.62\\
HD 179949       &    $-$4.66&    $-$4.76&    $-$4.72&     2.05&     0.68\\
HD 202206       &    $-$4.72&         &    $-$4.72&     2.04&     0.44\\
HD 210277       &    $-$5.02&    $-$5.07&    $-$5.06&     6.93&     2.25\\
HD 213240       &    $-$5.12&    $-$4.90&    $-$4.97&     5.11&     1.90\\
HD 216435       &    $-$4.95&    $-$5.00&    $-$4.98&     5.27&     1.56\\
HD 216437       &    $-$5.52&    $-$5.01&    $-$5.27&    12.96&     3.98\\
HD 217107       &    $-$5.17&    $-$5.05&    $-$5.08&     7.32&     1.40\\
HD 222582       &    $-$5.05&    $-$5.00&    $-$5.03&     6.16&     3.38\\
\hline
\end{tabular}
\label{cromoages}
\end{table*}

\begin{table*}
\center
\caption{Chromospheric index, Log R$'$$_{\rm HK}$, and age for the EH stars not observed
at the CASLEO}
\begin{tabular}{lcccclccc}
\hline
Name               &  $<$Log R$'$$_{\rm HK}$$>$  & D93 Age[Gy] & RPM98 Age[Gy] & & Name & $<$Log R$'$$_{\rm HK}$$>$
& D93 Age[Gy] & RPM98 Age[Gy]\\
\hline
16 Cyg B         &   $-$5.09 &       7.59&       3.79&   & HD 80606      &   $-$5.09 &       7.63&      1.73 \\
47 Uma           &   $-$5.02 &       6.03&        3.2&   & HD 88133      &   $-$5.16 &       9.56&      6.27 \\
51 Peg           &   $-$5.05 &        6.6&       2.21&   & HD 89744      &   $-$5.11 &       8.09&      2.55 \\
55 Cnc           &   $-$5.00 &        5.5&       1.21&   & HD 93083      &   $-$5.02 &          6&      3.86 \\
70 Vir           &   $-$5.07 &       7.09&       5.52&   & HD 99492      &   $-$4.94 &       4.49&      2.93 \\
BD$-$103166      &   $-$4.92 &       4.18&       0.53&   & HD 101930     &   $-$4.99 &       5.39&      3.48 \\
$\epsilon$ Eri   &   $-$4.46 &       0.66&       0.82&   & HD 102117     &   $-$5.03 &       6.21&      2.99 \\
$\gamma$ Cephei  &   $-$5.32 &      14.78&       6.39&   & HD 104985     &   $-$5.58 &      25.35&     27.08 \\
GJ 436           &   $-$5.21 &      11.05&       7.41&   & HD 106252     &   $-$4.97 &       5.02&      3.36 \\
GJ 876           &   $-$5.17 &        9.9&       6.52&   & HD 108874     &   $-$5.08 &       7.26&      2.21 \\
GJ 777A          &   $-$5.07 &       7.09&       2.08&   & HD 111232     &   $-$4.98 &        5.2&      9.65 \\
HD 3651          &   $-$4.98 &       5.13&       2.25&   & HD 117618     &   $-$4.90 &       3.88&      2.72 \\
HD 4203          &   $-$5.16 &       9.41&       1.66&   & HD 128311     &   $-$4.40 &       0.39&      0.41 \\
HD 8574          &   $-$5.07 &       7.13&       3.79&   & HD 136118     &   $-$4.93 &       4.26&      3.16 \\
HD 8673          &   $-$4.71 &       1.95&       0.01&   & HD 145675     &   $-$5.09 &        7.6&      1.20 \\
HD 10697         &   $-$5.12 &       8.48&       3.49&   & HD 150706     &   $-$4.57 &       1.17&      0.83 \\
HD 11964         &   $-$5.16 &       9.56&       6.27&   & HD 154857     &   $-$5.14 &       8.98&     14.29 \\
HD 12661         &   $-$5.07 &       7.05&       1.39&   & HD 177830     &   $-$5.35 &      15.89&      4.03 \\
HD 16141         &   $-$5.09 &       7.76&       3.08&   & HD 178911 B   &   $-$4.98 &        5.2&      1.54 \\
HD 20367         &   $-$4.50 &       0.87&       0.37&   & HD 187123     &   $-$4.99 &       5.33&      2.26 \\
HD 23596         &   $-$5.06 &       6.89&       1.61&   & HD 190228     &   $-$5.18 &      10.16&     14.29 \\
HD 37124         &   $-$4.86 &       3.33&       6.68&   & HD 192263     &   $-$4.44 &       0.57&      0.55 \\
HD 40979         &   $-$4.63 &       1.48&       0.51&   & HD 195019     &   $-$4.99 &       5.33&      2.58 \\
HD 41004A        &   $-$4.66 &       1.64&       1.48&   & HD 196050     &   $-$4.85 &       3.17&      1.03 \\
HD 45350         &   $-$5.00 &       5.59&       1.94&   & HD 208487     &   $-$4.90 &       3.88&      4.06 \\
HD 46375         &   $-$4.97 &       4.96&       1.68&   & HD 209458     &   $-$4.95 &       4.72&      2.88 \\
HD 49674         &   $-$4.77 &       2.38&       0.55&   & HD 216770     &   $-$4.88 &        3.6&      1.02 \\
HD 50554         &   $-$4.95 &       4.58&       2.89&   & HD 219449     &   $-$5.47 &      20.76&     18.24 \\
HD 50499         &   $-$5.02 &          6&       1.82&   & HD 330075     &   $-$5.03 &       6.21&      3.09 \\
HD 68988         &   $-$5.06 &       6.78&       1.34&   & $\rho$ Crb    &   $-$5.06 &       6.94&      8.48 \\
HD 70642         &   $-$4.90 &       3.88&       1.42&   & $\tau$ Boo    &   $-$4.78 &       2.52&      0.80 \\
HD 73256         &   $-$4.49 &       0.83&       0.26&   & TrES$-$1      &   $-$4.77 &       2.41&      1.63 \\
HD 74156         &   $-$5.08 &       7.38&       2.83&   & $\upsilon$ And&   $-$4.99 &       5.32&      2.26 \\
\hline
\end{tabular}
\label{cromoagesnotcasleo}
\end{table*}

The uncertainty in the ages derived by the CE method strongly depends
on how well the activity cycle for a particular object has been monitored
\citep[see, for example,][]{donahue}.
For example, \cite{henry2000} estimated that if the star happens to be in a
phase similar to the Solar Maunder minimum, which can last for several decades,
the CE age estimation can be overestimated by $\sim$ 2--5 Gyr.
However, if the star is in a ''maximum'' phase of the activity cycle
the uncertainty in the age may be smaller. \cite{hen} noted
that the D93 relation yields ages such as in 15 out of 22 binaries
the ages differ by less than 0.5 Gyr.  In general, \cite{gus} has
estimated a typical uncertainty in the ages derived by the CE method
of roughly 30\%.

The sample of EH stars has at least 19 multiple systems \citep{mult}, including
three close binaries, \hbox{$\gamma$ Cep} \citep{R51}, \hbox{HD 41004} A and B
\citep{mult1,mult3}, and \hbox{GJ 86} \citep{mult2}.
Assuming that the binary components are coeval,
\cite{donahue} found that the age discrepancy between
both stars has the same order as the uncertainty
in the chromospheric age derivation itself.
He found that for stars older than 2 Gyr, the
age uncertainty is typically below 1 Gyr.
This difference is, then, probably due to non-synchronized phases
in the activity cycle at which each individual star has
been monitored.  Tidal interactions may, in principle, affect the
stellar activity in close systems. However, the analyzed sample includes
a relatively small number of these type of binaries and thus this effect
cannot significantly alter our statistical results.

An enhancement in the CE due to the presence of a close giant
planetary companion has been investigated by several authors
\citep[see, for example,][]{cuntz,sacu,shkolnik}.  Moreover,
\cite{rs} suggested that close giant planets may stimulate
the presence of ''superflares'' on the CE of the EH stars.
\cite{san05} proposed that the photometric variability
observed in HD 192263 may also be related to the
star-planet interaction effect. Other EH stars
have been searched for an enhancement in the CE
due to the presence of a close planet \citep{sacu,shkolnik04}.

\cite{shkolnik} found evidence for a planet-induced chromospheric
activity in the EH star HD 179949, having a planetary companion
with a semi-major axis of $\sim$ 0.04 AU (an orbital period of $\sim$ 3 days).
The planet period is synchronized with the enhancement of the CE,
which increases by $\sim$ 4\% when the planet passes in front of the star.
Translated into ages, this would represent a difference of 0.8 Gyr for a
5 Gyr EH star, adopting the D93 calibration. This difference is about the
same as the uncertainty in the chromospheric ages.  A similar effect was
detected in $\nu$ And \citep{shkolnik,shkolnik04}.

\begin{figure}
\center
\includegraphics[width=60mm]{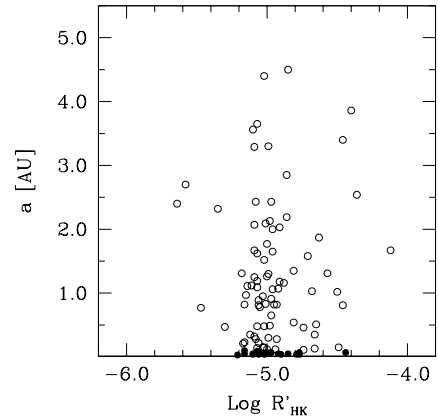}
\caption{CE (measured by Log R$'$$_{\rm HK}$) versus the semi-major axis a, for
the EH sample. 51 Peg-like objects (i.e., those with a $<$ 0.1 AU) and the
rest of the sample are represented by filled and empty circles, respectively.}
\label{logra}
\end{figure}

We have searched for correlations between the CE (measured
by Log R$'$$_{\rm HK}$) and the orbital parameters of the
associated planet, such as: M~ sin~ {\it i}, e, and a.
Figure \ref{logra} shows the Log R$'$$_{\rm HK}$ vs
the semi-major axis, a,  plot, as an example. In this figure,
51 Peg-like stars (i.e., those with a $<$ 0.1 AU) are indicated
with filled circles whereas the rest of the sample is with empty symbols.
In general, no clear trend is found between the CE and the planet orbital
parameters.  However, the chromospheric ages may be affected,
particularly in the cases of HD 179949, HD 192263,
and $\nu$ And. In any event, we expect that the CE enhancement, due
to the presence of a close giant planet, to be in about the same order
as the uncertainty in the chromospheric ages in view of the amount
of this effect for HD 179949, as discussed above.

\begin{figure}
\center
\includegraphics[width=60mm]{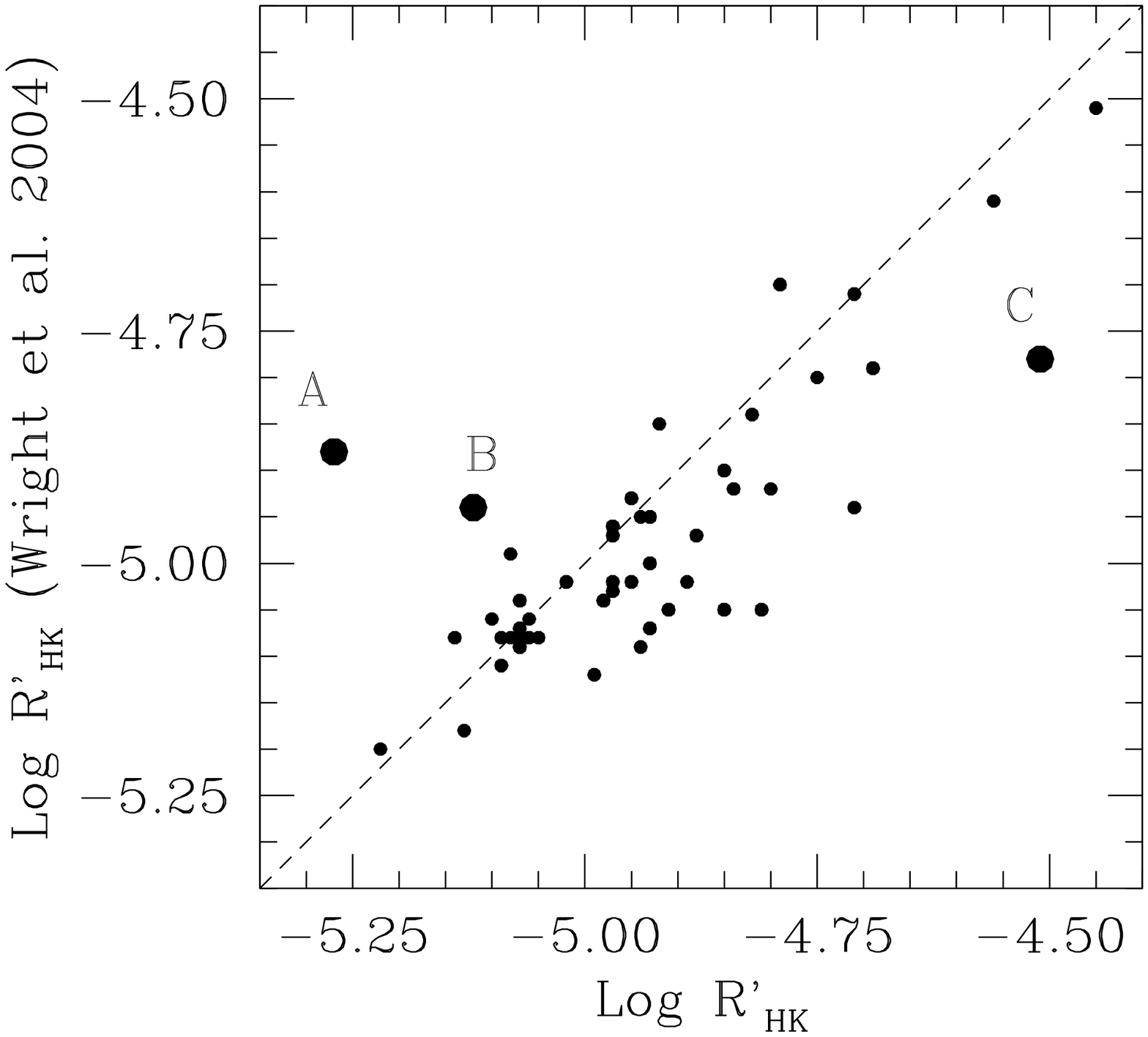}
\includegraphics[width=60mm]{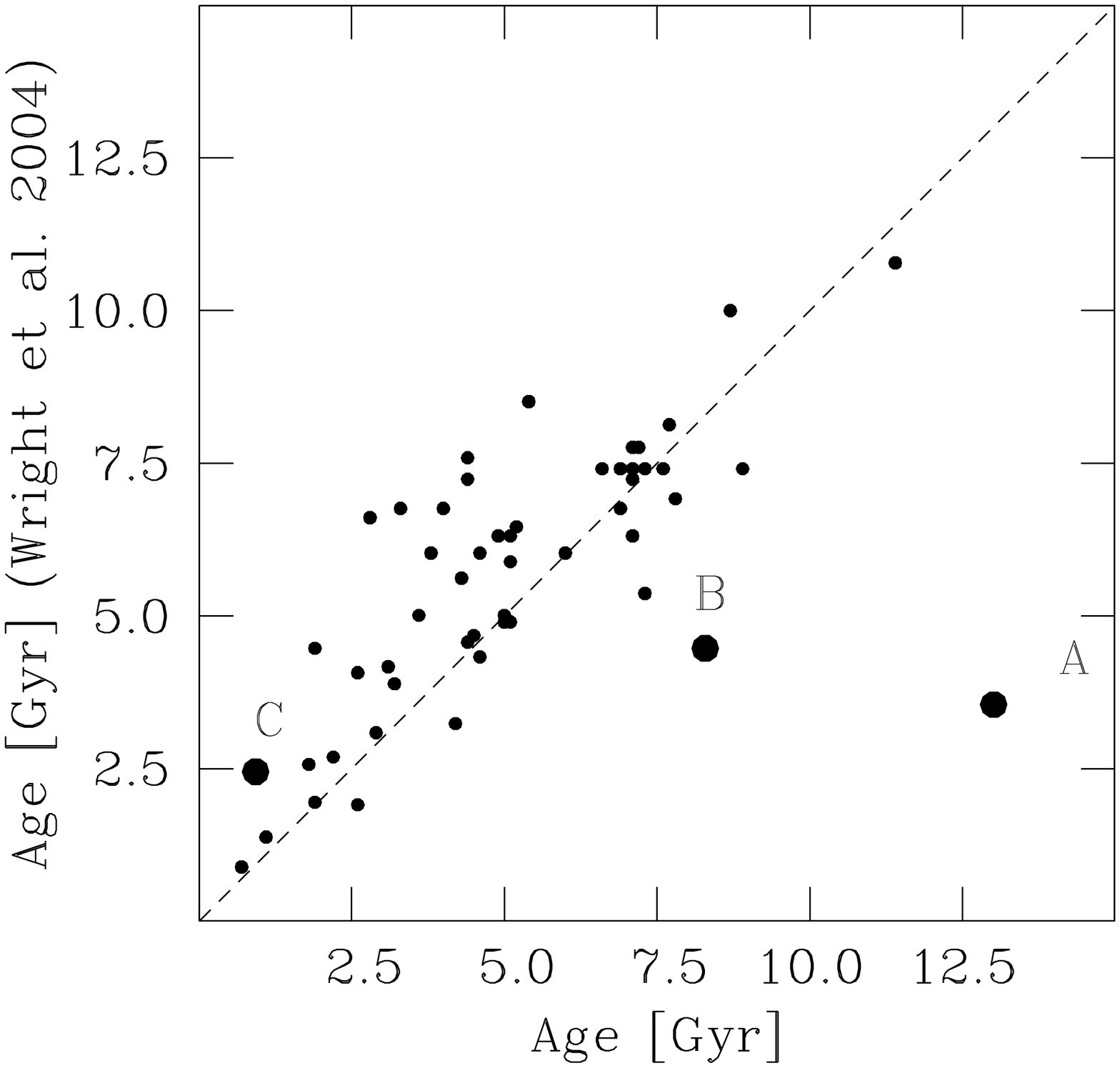}
\caption{Comparison between the CE index, Log R$'$$_{\rm HK}$, and the age
for the H sample stars derived by \cite{wright2} and those
reported in this contribution. To make these comparisons we have
used stars in common and eliminated \cite{wright2}'s data from the averages
reported in Table \ref{cromoref}.  The ages have been obtained from the
D93 calibration. The large circles indicate stars with the largest discrepancies:
(A) HD 19994, (B) HD 89744 and (C) HD 130322, see Table \ref{diff}.}
\label{w1w2}
\end{figure}

Recently, \cite{wright2} have derived the CE for a sample of $\sim$ 1200 F-, G-, K-
and \hbox{M- type} main-sequence stars, using archival spectra from the California \&
Carnegie Planet Search Project.
To somehow compensate or smooth the effect of the stellar variability in
an uneven sampling set of data, they used the median S-values in
30-day bins and then adopted the median value of those medians (the
''grand-S'' value).  The number of interval-bins typically vary between
a few and few tenths, during a period of time of more than $\sim$ 6 years.
These authors derived ages by applying
the D93 calibration.  As \cite{wright2}'s sample includes 63 EH
stars, in Figure~\ref{w1w2} we compare their values of Log R$'$$_{\rm HK}$
and ages with the ones reported in this paper, excluding, in this case,
\cite{wright2}'s data from our compilation (see Table \ref{cromoref}).

We notice a general agreement between the CE indexes (and the corresponding
ages) derived by \cite{wright2} and our compilation. The median difference in
Log R$'$$_{\rm HK}$ is 0.04 dex (with a standard deviation of 0.05 dex),
implying a median age discrepancy of $\sim$ 0.4 Gyr (\hbox{0.8 Gyr} for the standard
deviation) for a 5 Gyr star. The largest CE (and age) differences correspond to the
stars HD 19994, HD 89744 and HD 130322, see Table \ref{diff}. These objects
are marked in \hbox{Figure \ref{w1w2}} with the letters A, B and C, respectively.

\begin{table*}
\center
\caption{EH stars with the largest CE and age differences}
\begin{tabular}{lccrcc}
\hline
Name & Log R$'$$_{\rm HK}$ & Log R$'$$_{\rm HK}$ & Age [Gyr] & Age [Gyr] & Label in Figure \ref{w1w2} \\
     & this paper & \cite{wright2} & this paper & \cite{wright2} \\
\hline
HD 19994  &  $-$5.27 & $-$4.88 & 13.01  & 3.55 & A \\
HD 89744  &  $-$5.12 & $-$4.94 &  8.29  & 4.47 & B \\
HD 130322 &  $-$4.51 & $-$4.78 &  0.93  & 2.45 & C \\
\hline
\end{tabular}
\label{diff}
\end{table*}

\section{Derivation of ages for the EH stars applying other methods}

In this section we apply four other techniques to infer ages
for the EH group and compare these results with the chromospheric
determinations.

\subsection{Isochrones}

The T$_{\rm eff}$ and the luminosity of a star allow us to place it
on the theoretical HR diagram. This position changes as the star evolves.
The age of a given star can, at least in principle, be inferred
adopting an evolutionary model and the corresponding isochrones.
However, in practice, the derivation of reliable isochrone ages
is a difficult task.

Isochrone ages are usually calculated by comparing the position of
the star on the HR diagram with the set of isochrones adopted. This
procedure is particularly complicated for low mass objects
for which isochrones are curved and approximate one to the other.
Apart from this, the uncertainties in the observables (i.e., Teff, luminosity
and metallicity) can seriously limited the applicability
of this technique. \citet{pont} have extensively discussed
the influence on the age derivation of these uncertainties.
In addition these authors have developed a
method based on Bayesian probability to treat systematic
bias and large uncertainties in the observables to derive
more reliable ages. In particular the use of this method
allowed a considerable reduction  the intrinsic dispersion
in the [Fe/H]-age relation \citep[see][]{pont}.

Recently \cite{nord} have determined
ages for about 13636 nearby stars, including the EH group, using the Padova
isochrones \citep[][]{gir,salas}. This set of isochrones covers the
range of ages between \hbox{0 and 17.8 Gyr}.
\cite{nord} give the age for a given star if it lies within 1$\sigma$ upper and lower
limits of the nearest isochrone trace. Otherwise, only upper or lower limits
are determined.  In Table \ref{allages}, columns 2, 3 and 4, we list the ages
and/or lower and upper limits derived by these authors for the EH group.
As mentioned before, our purpose is to compare the ages derived by applying
different methods. The compilation of ages from \cite{nord} in Table \ref{allages}
is useful to this purpose.

\cite{nord} have provided estimations of errors in their age
determinations. As mentioned above their sample includes 13636 stars, with
84\% of the objects lying within 1$\sigma$ upper and lower
limits. In addition, 82\% of these stars has estimated relative
errors below 50\%, and 47\% of these even below 25\%.
We follow \cite{nord} to quote errors for the isochrone
ages for the EH stars. 77\% (61 out of 79) of the
EH sample lies within 1$\sigma$ upper and lower limits.
44\% of these objects (27 out of 61) has estimated relative
errors below 50\% and 10\% of these (6 out of 61) even below
25\%. On average, we assume a ''typical'' error of $\sim$ 50\%
in the isochrone age determinations.

We note that \cite{nord} carried out a careful and detailed
error estimation for the isochrone technique (the reader is
referred to their work for the description of the method applied).
In general, these authors derived larger uncertainties than
those classical ones obtained \cite[see, for example,][]{gus,edv}.
However, \cite{nord}'s error analysis seems more realistic
than those performed before. In addition, these authors have
also compared their age derivations with those computed using the Bayesian
method of \citet{pont}, finding no significant differences due to
the procedures themselves.

\subsection{Lithium abundance}

The lithium content in the stellar atmosphere is destroyed as
the convective motions gradually mix the stellar envelope
with the hotter (T $\sim$ 2.5 $\times$ 10$^6$ K)
inner regions. Thus a lithium-age relation may be expected.
However, this relation is poorly constraint.
For instance, while \cite{boes91} derived a lithium-age relation for stars with
5950 K $<$ T$_{\rm eff}$ $<$ 6350 K belonging to eight open clusters,
\cite{pasq2} and \cite{pasq1} found a factor of 10 for the lithium abundance for stars
at given T$_{\rm eff}$ in M67, an open cluster with almost the solar age and metallicity.

The lithium abundance of EH and field stars has been compared in the
literature.  \cite{gl} suggested that the EH stars have less lithium
than field stars, while \cite{ryan} proposed that both groups have similar lithium
abundance.  Recently, \cite{isra} found a likely excess of lithium depletion in
EH stars with T$_{\rm eff}$ in the range 5600--5850 K, in comparison to
field stars in the same range of temperatures. For an average difference
of \hbox{$\sim$ 1.0 dex} in the lithium content of the two groups, the EH stars are
$\sim$ 2 Gyr older than field stars, according to the lithium-age relation derived by
\cite{soder2}. On the other hand, \cite{isra} did not find a significant
difference for EH and field stars with T$_{\rm eff}$ in the
range 5850--6350 K.

We apply this method only as an additional age indicator to compare with
the other techniques.  We obtained lithium abundances, Log N(Li), for the sample
EH stars from \cite{isra} and used the calibration of \cite{soder2} to
derive ages for these objects.  This relation is valid for solar-type stars with
abundances intermediate to the Hyades and the Sun
(i.e., \hbox{0.95 $<$ Log N(Li) $<$ 2.47)}. 20 EH stars are included
within these limits. This range in the Li abundances introduces
a bias toward younger ages and prevents us from making a meaningful
comparison with the other estimators. We then use the \cite{soder2}
calibration to derive individual stellar ages for the EH stars when having Li abundances
within the valid range.  In Table \ref{allages} column 5 gives the obtained
ages. For HD 12661 only an upper limit to the lithium abundance was
available and thus we derived a lower limit to the age of this object.
The subindex ''L'' in Table \ref{allages} indicates so.

\citet{boes91} derived a lithium-age calibration valid over a larger
rage in Li abundances, \hbox{2.1 $<$ Log N(Li) $<$ 3.0}, but
restricted to stars with \hbox{5950 K $<$ T$_{\rm eff}$ $<$ 6350 K}.
This temperature range comprises \hbox{$\sim$ 20\%} of the EH sample.
In addition, considering both the Li abundances and the T$_{\rm eff}$
intervals, the \citet{boes91}'s calibration can be applied to
obtain ages for only 9 EH stars. For this reason we chose the
\citet{soder2}'s calibration over the \citet{boes91}'s relation.
We note that these calibrations are complementary in metallicity
range, however they do not provide a reasonably good agreement
within the metallicity interval held in common, thus preventing the
application of both calibrations together in our age estimations.

\subsection{Metallicity}

The production of heavy elements in the stellar cores during the life time of
the Galaxy enriches the interstellar medium from which new stars are formed. Thus
an age-metallicity relation may be expected. \cite{twarog} and \cite{edv} studied
the disk population and found a relatively weak correlation between these two quantities.
The scattering in metallicity for a given age is so large that some
authors have even questioned the existence of a correlation between these two
parameters \citep[see, for example,][]{felt,ibuki,nord}.
Other works have been aimed to improve this relation, trying to
disentangle the different possible contributions to the dispersion \citep[e.g.,][]{ngbert,reddy}.
In particular, \citet{pont} have shown that at least part of the scattering in
the original \cite{edv}'s age-metallicity relation is mainly due to systematic
bias affecting the ages derived by the isochrones method. However, in spite
of these efforts, the dispersion at each given age is still rather large. Nevertheless,
the use of the age-metallicity relation as an independent age estimator can
provide some constrains to the EH stars age distribution and a manner to check the
results derived by the chomospheric method.

We obtained the spectroscopic [Fe/H] data for the sample of EH stars from
\cite{santos} whenever possible, otherwise we used the data from \cite{nord}.
We adopted the age-metallicity relation of \cite{carr} and followed the procedure
outlined by \cite{lacha} only to derive upper limits to the ages of the EH stars,
due to the large scattering in this relation.

We defined an upper envelope to the data points in the \cite{carr} relation,
binning these data in 3 Gyr intervals and calculating the average [Fe/H] and
the corresponding dispersion.
We adopted as the upper limit the average [Fe/H] plus
the rms values in each bin.
We then fitted these points with a quadratic polynomial by least squares,
giving t$_{\rm max}$ (an upper limit to the age) as function of the metallicity.
Table \ref{fefit} lists the data derived from \cite{carr} used to calculate the
following relation:
\vskip 0.1in

\noindent
\begin{equation}
{\rm t_{max} = -35.847~~~[Fe/H]^{2} - 31.172~~~[Fe/H] + 6.9572.}
\label{seis}
\end{equation}
\vskip 0.1in

\begin{table}
\center
\caption{Data points used to calculate Equation \ref{seis} derived from \cite{carr}}
\begin{tabular}{rr}
\hline
[Fe/H]   & t$_{\rm max}$ [Gyr] \\
\hline
    0.15    & 1.5   \\
    0.07    & 4.5   \\
$-$ 0.02    & 7.5   \\
$-$ 0.14    & 10.5   \\
$-$ 0.35    & 13.5   \\
\hline
\end{tabular}
\label{fefit}
\end{table}

\noindent
Equation \ref{seis} is valid for $-$0.35 $<$ [Fe/H] $<$ 0.15 or
1.5 $<$ t$_{\rm max}$ $<$ 13.5 Gyr. This range of metallicities
is appropriate for 44\% (50 out of 114) of the EH objects, for which it was
possible to apply this age estimator. This range excludes the most metal-rich EH stars,
introducing a bias toward older ages. Table \ref{allages} in
column 6 lists metallicity derived ages for the individual objects
comprised within the before mentioned range of metallicity. \citet{carr} data
can also be used to derive a lower metallicity-age limit. The resulting
relation includes only 11\% (13 of out 114) of the EH objects, due
to their metal-rich nature.

\subsection{Kinematics}

The velocity dispersions of a coeval group of stars increases with time
\citep[][]{kin1,kin2,kin4,kin3}. The kinematics of a given group can be used as an
age estimator for the group rather than to derive individual stellar ages.
In particular, the transverse velocity dispersion S (i.e., the dispersion of the velocity
components perpendicular to the line of sight) of a sample of stars
may be used with this purpose \citep{kin3}.

We derived S for a sample of solar neighborhood stars, to compare with the
EH group. The first sample was constructed in the same manner described
by \cite{kin4}. These authors selected a kinematically unbiased solar neighborhood
group isolating a magnitude-limited subsample, composed by single main sequence
stars with relative parallax errors smaller than 10\% obtained from the
Hipparcos catalog. In what follows we briefly outline \cite{kin4}'s procedure
to define such a sample.

Due to the lack of completeness in the Hipparcos catalog\footnote{See \cite{kin4}
for details on this issue.}, \cite{kin4} used
the Tycho catalog to construct a combine sample that they estimated to be
95\% complete. The Hipparcos catalog was divided in 16$\times$16$\times$10
uniformly spaced bins in sin $b$, Galactic longitude $l$ and B$-$V.  All stars
brighter in magnitude than the second brightest star per bin included in Tycho
and not in the Hipparcos catalog were added to this bin. In our case this
gave a list of 8864 stars.

A second sample was constructed from the Hipparcos Proposal 018, containing
6845 stars within 80 pc and south of $-$28 deg,
which have been spectrally classified by the Michigan Catalog
by 1982 \citep{mich1,mich2,mich3}. Of these stars, 3197 are main-sequence
single objects with relative parallax errors smaller than 10\%.  The union
of these two groups provides a kinematically
unbiased sample of 12061 stars \citep[see also][]{kin4}.

\begin{figure}
\center
\includegraphics[width=80mm]{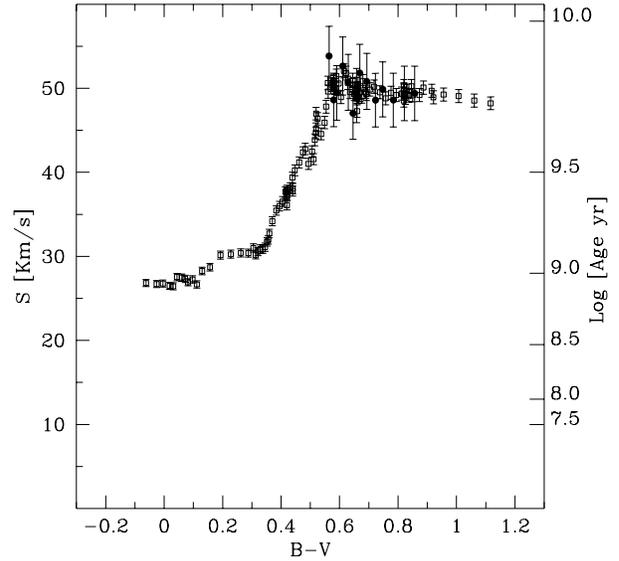}
\caption{Transverse velocity dispersion S vs dereddened B$-$V. Solar neighborhood
stars and EH stars are represented by squares and filled circles, respectively.
Error bars are derived as $\Delta$S $=$ S$/\sqrt{2{\rm n}-2}$, where n is the number of objects
per bin. The age scale on the right side was obtained from \cite{kin3}.}
\label{kinemvega}
\end{figure}

The velocity dispersion S for each star was derived following
the formalism of \cite{kin4}. Figure \ref{kinemvega} shows the
S vs B$-$V diagram for the solar neighborhood with open squares.
B$-$V colors were also obtained from the Hipparcos catalog and
dereddened according to the spectral types, given in the same catalog.
We used a sliding window of 500 objects and plotted a
point every time a 100 stars are left out from the window.
We have tested different sizes for the sliding window
and found no significant differences. The global change in the
slope at B$-$V $\sim$ 0.6 for solar neighborhood stars
in Figure \ref{kinemvega} is called Parenago's
discontinuity, and it has recently been quantified by
\cite{kin4}.  At the red side of this point, stars of every
age are found, while at the blue side only the most recently
formed objects lie. The discontinuity itself corresponds
to the B$-$V color at which the main sequence lifetime of a
star equals the age of the Galactic disk \citep{kin4}.
The general trend of the solar neighborhood sample
in Figures \ref{kinemvega} is very similar to that derived by \cite{kin3}.

We obtained S for the EH sample including stars with
relative parallax error less than 10\%.
This excludes 7 out of 131 EH stars with parallax data
(i.e., 5\% of the group). However we were unable to
apply the single and main sequence stars criteria as
these would eliminate $\sim$ 42\% of the objects.
In Figure \ref{kinemvega}, with filled circles,
we superimpose the results
for the EH stars to compare with the solar
neighborhood sample. For the EH group we chose a
sliding window of 30 objects and plotted a point when
6 stars have left the window. On the right side
of Figure \ref{kinemvega} we indicate the
age scale derived by \cite{kin3}.

EH stars seem to have similar transverse velocity dispersions to the
solar neighborhood stars, with an average age of about 4--6 Gyr.
However, most EH stars lie to the right of Parenago's discontinuity,
where the kinematic method becomes less reliable.

\cite{kin5} analyzed the sample of Vega-like candidate stars,
(main sequence) stars that show infrared excesses, attributed to
the presence of circumstellar dust (T $\sim$ 50--125 K), warmed by the
central object \citep[e.g.,][]{zuck}. They found that the
transverse velocity dispersions are systematically smaller than for solar
neighborhood stars, suggesting a younger age for the
Vega-like sample with respect to the solar neighborhood.
As most of the Vega-like stars have A spectral type,
the sample lies to the left of Parenago's discontinuity,
where the kinematic method can safely be applied.

As an additional test we have used the space velocity components (U, V, W)
to check the consistency of the results. \cite{reid} has recently applied this
technique to derive a lower limit to the age of a group of 67 EH stars.
We extend this analysis to include 101 of the presently known EH stars
with space velocity data available.
We followed the kinematic method described by \cite{lacha}, using values of U, V
and W for the EH stars from \cite{nord} and correcting them by the Solar Motion
\citep[$-$13.4, $-$11.1, $+$6.9 Km/s;][]{chen}.
We derived a lower limit of 3.9 $\pm$ 2.9 Gyr for the sample of EH stars,
which agrees with the average age derived using the transverse velocity dispersions
(i.e., 4--6 Gyr).  We caution, however, that the later kinematic method is less reliable
as an age estimator because EH stars lie red-ward of Parenago's discontinuity.

The kinematic properties of the EH stars have also been investigated
by other authors. As mentioned before, \citet{reid} used the space
velocity components to infer a lower limit to the age for the
EH sample. Other groups have confronted kinematic and metallicity
properties of the EH stars with those of similar stars with non known planets.
The aim of these works was to analyze whether the kinematics of the EH stars
may provide a hint to explain the relatively high metallicities of the EH
stars.

For example, \citet{gon99} analyzed the chemical and dynamical
properties of the EH stars within the framework of the diffusion of
stellar orbits in space \citep{w96}.  With this purpose he
confronted the EH group which is metal rich with a sample of F and G
dwarfs and subgiants with well determined metallicities, ages and kinematics. He
concluded that the stellar diffusion model is not able to account for the
high metallicity in EH stars.

\citet{bargra03} compared the galactic orbits of
EH stars with those of stars with no known planets taken from
the catalog of \citet{edv}. Both groups are not
kinematically different. However at each perigalactic distance
the EH stars have systematic larger metallicities than the
average of the comparison sample.  This result was recently
confirmed by \citet{laws03}. The latter authors also
found evidence of a difference in the slope of the metallicity-
Galactocentric radius relation between star with and without
planets. Due to the metal rich nature of the EH stars, these
objects have a steeper slope than the comparison group.

\citet{san03b} analyzed the space velocities of EH stars and
stars with no planet/s detected, in relation to their different
metallicity abundances. They concluded that the space velocity distribution
for the first objects is basically the same as for the second type
of stars. However the EH stars lie on the metal-rich envelope of the
no-known-planet-star population.

\begin{table*}
\center
\caption{Ages derived from isochrone, lithium and [Fe/H] abundances. ``L" indicates
a lower limit.}
\begin{tabular}{lrrrlcr}
\hline
                & Isochr.  & Isochr.  & Isochr.  &  Lithium  & [Fe/H]       \\
                & Age      & Min. Age & Max. Age &  Age      & Max. Age     \\
Object          &  [Gyr]   & [Gyr]    & [Gyr]    & [Gyr]     & [Gyr]        \\
\hline
16 Cyg B        &      9.9&      5.6&     13.2&         &      4.2     \\
47 Uma          &      8.7&      5.3&     11.9&      2.0&      5.0     \\
51 Peg          &      9.2&      4.8&     12.0&      3.5&              \\
70 Vir          &      7.4&      6.7&      7.9&      1.9&      8.7     \\
$\epsilon$ Eri  &         &         &         &         &     10.4     \\
GJ 86           &         &         &         &         &     12.5     \\
Hip 75458       &         &         &         &         &      2.3     \\
$\rho$ Crb      &     12.1&     10.1&     13.9&      3.3&     11.9     \\
$\tau$ Boo      &      2.4&      1.3&      3.1&         &              \\
$\upsilon$ And  &      3.3&      2.8&      5.0&         &      2.3     \\
HD 142          &      3.6&      2.8&      4.3&         &      1.9     \\
HD 1237         &         &      8.8&         &         &      2.7     \\
HD 2039         &      1.8&         &      3.4&         &              \\
HD 3651         &     17.0&      2.6&         &         &      2.7     \\
HD 4208         &         &         &         &         &     12.4     \\
HD 6434         &     13.3&      7.0&         &         &              \\
HD 8574         &      8.2&      5.7&      9.6&         &      5.0     \\
HD 8673         &      2.8&      2.1&      3.3&         &      8.7     \\
HD 10647        &      4.8&         &      7.0&         &      7.9     \\
HD 10697        &      7.1&      6.4&      7.9&      1.5&      1.9     \\
HD 12661        &         &         &         &      4.4&              \\
HD 16141        &     11.2&      9.7&     12.9&      4.0&              \\
HD 17051        &      3.6&      1.1&      6.7&         &              \\
HD 19994        &      4.7&      3.1&      5.2&      1.4&              \\
HD 20367        &      6.4&      3.6&      8.9&         &              \\
HD 23079        &      8.4&      5.3&     12.6&         &     10.0     \\
HD 23596        &      5.4&      3.1&      6.7&         &              \\
HD 28185        &     12.2&      7.1&         &      2.8&              \\
HD 33636        &      8.1&      0.1&     13.4&         &      9.2     \\
HD 34445        &      9.5&        8&     11.1&         &      7.3     \\
HD 39091        &      6.0&      2.9&      9.0&         &      3.5     \\
HD 40979        &      6.2&      3.8&      9.2&         &              \\
HD 41004A       &         &         &         &         &      9.5     \\
HD 45350        &     12.6&     10.4&     14.6&         &              \\
HD 46375        &     16.4&      7.7&         &         &              \\
HD 50554        &      7.0&      3.3&      9.9&         &      6.6     \\
HD 50499        &      4.3&      2.8&      7.4&         &              \\
HD 52265        &      3.8&      1.6&      5.2&         &              \\
HD 65216        &         &         &         &         &     10.2     \\
HD 68988        &      3.7&      1.5&      6.1&         &              \\
HD 70642        &     10.2&      4.2&     16.0&         &              \\
HD 72659        &      8.2&      6.5&      9.6&         &      6.0     \\
HD 73256        &     15.9&      6.4&         &         &              \\
HD 73526        &     10.3&      8.3&     12.6&         &              \\
HD 74156        &      3.2&      2.7&      3.7&         &              \\
HD 75289        &      4.0&      2.2&      5.8&         &              \\
HD 76700        &     11.5&     10.0&     13.1&         &              \\
HD 80606        &         &         &     17.6&         &              \\
HD 82943        &      3.5&         &      7.2&         &              \\
\hline
\end{tabular}
\end{table*}

\setcounter{table}{6}
\begin{table*}
\center
\caption{Continued.}
\begin{tabular}{lrrrlcr}
\hline
                & Isochr.  & Isochr.  & Isochr.  &  Lithium  & [Fe/H]       \\
                & Age      & Min. Age & Max. Age &  Age      & Max. Age     \\
Object          &  [Gyr]   & [Gyr]    & [Gyr]    & [Gyr]     & [Gyr]        \\
\hline
HD 89307        &      8.8&      3.9&       13&         &     12.2     \\
HD 89744        &      2.2&      2.0&      2.4&      1.2&              \\
HD 92788        &      9.6&      4.8&     14.3&      3.4&              \\
HD 102117       &     12.6&     10.9&     14.3&         &      3.9     \\
HD 104985       &      3.1&      2.3&      3.6&         &              \\
HD 106252       &      9.2&      5.2&     13.5&      2.5&      7.3     \\
HD 108147       &      4.4&      2.3&      6.6&         &              \\
HD 108874       &     14.1&     10.7&         &         &              \\
HD 111232       &         &      8.9&         &         &              \\
HD 114386       &         &         &         &         &      9.2     \\
HD 114729       &     11.9&     10.4&     13.3&      1.6&     12.5     \\
HD 114762       &     11.8&      7.9&     15.1&         &              \\
HD 114783       &         &         &         &         &      3.9     \\
HD 117207       &     16.1&     11.1&         &         &      4.6     \\
HD 117618       &      6.7&      3.6&      9.6&         &      7.6     \\
HD 121504       &      7.1&      3.9&     10.2&         &              \\
HD 128311       &         &         &         &         &      6.0     \\
HD 130322       &         &         &         &         &      6.0     \\
HD 134987       &     11.1&      6.4&     12.6&         &              \\
HD 136118       &      4.8&      2.9&      5.5&         &      8.1     \\
HD 141937       &      1.8&         &      7.5&         &      3.5     \\
HD 142022       &     17.2&      9.4&         &         &              \\
HD 142415       &      2.4&         &      7.9&         &              \\
HD 147513       &      8.5&         &     14.5&      1.3&      5.0     \\
HD 150706       &      8.0&         &     15.0&         &      7.3     \\
HD 154857       &      3.5&      2.9&      4.4&         &     13.1     \\
HD 162020       &         &         &         &         &      9.5     \\
HD 168443       &     10.6&      9.5&     11.8&         &      5.0     \\
HD 168746       &     16.0&     10.8&         &         &      9.2     \\
HD 169830       &      2.3&      1.9&      2.7&      3.9&              \\
HD 179949       &      3.3&      0.4&      5.4&         &              \\
HD 183263       &      3.3&      1.1&      5.8&         &              \\
HD 187123       &      7.3&      3.8&     10.6&      3.8&      2.3     \\
HD 188015       &     10.8&        6&         &         &              \\
HD 190228       &      5.1&      4.1&      6.6&      3.7&     12.5     \\
HD 192263       &         &         &         &         &      7.6     \\
HD 195019       &     10.6&      9.4&     11.8&      3.0&      3.9     \\
HD 196050       &      3.5&      1.8&      5.3&         &              \\
HD 196885       &      8.4&      7.2&      9.7&         &              \\
HD 202206       &         &         &         &      4.2&              \\
HD 208487       &      6.6&      3.8&      9.3&         &     10.8     \\
HD 209458       &      6.6&      3.5&      9.2&         &      6.3     \\
HD 213240       &      3.6&      3.0&      4.2&         &              \\
HD 216435       &      5.4&      4.9&      6.0&         &              \\
HD 216437       &      8.7&      7.5&      9.7&      1.6&              \\
HD 216770       &     16.9&      7.4&         &         &              \\
HD 217107       &         &      6.5&         &         &              \\
HD 222582       &     11.1&      6.9&     15.3&         &      5.3     \\
HD 330075       &         &         &         &         &      4.2     \\
\hline
\end{tabular}
\label{allages}
\end{table*}

\subsection{Comparison of ages for the EH stars derived by different methods}

In this section we compare the age distributions of the EH stars
derived by the chromospheric, isochrone, Li and [Fe/H] abundances.
The kinematic ages are not considered here as non individual stellar
ages can be obtained.

\begin{figure}
\center
\includegraphics[width=60mm]{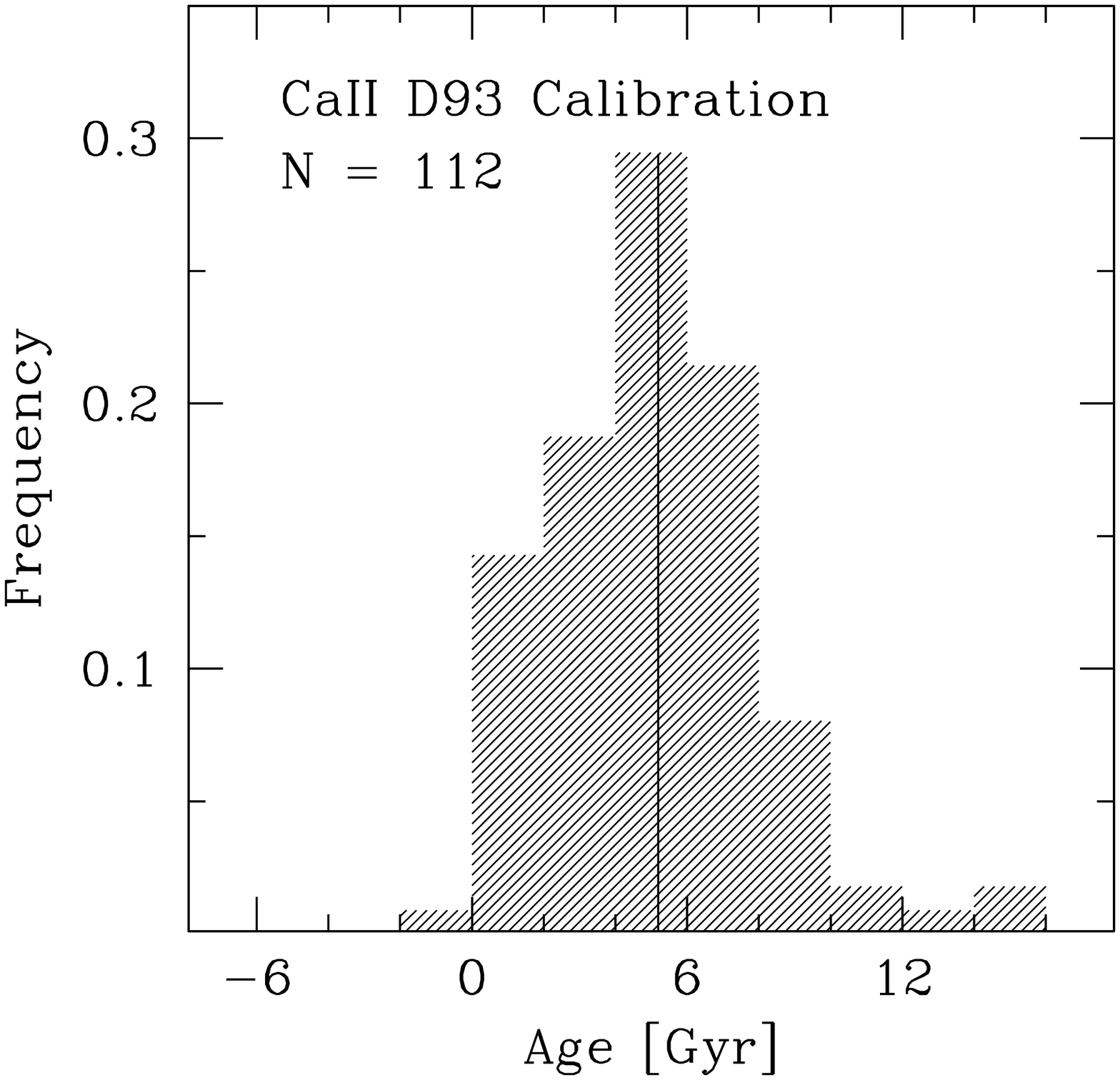}
\includegraphics[width=60mm]{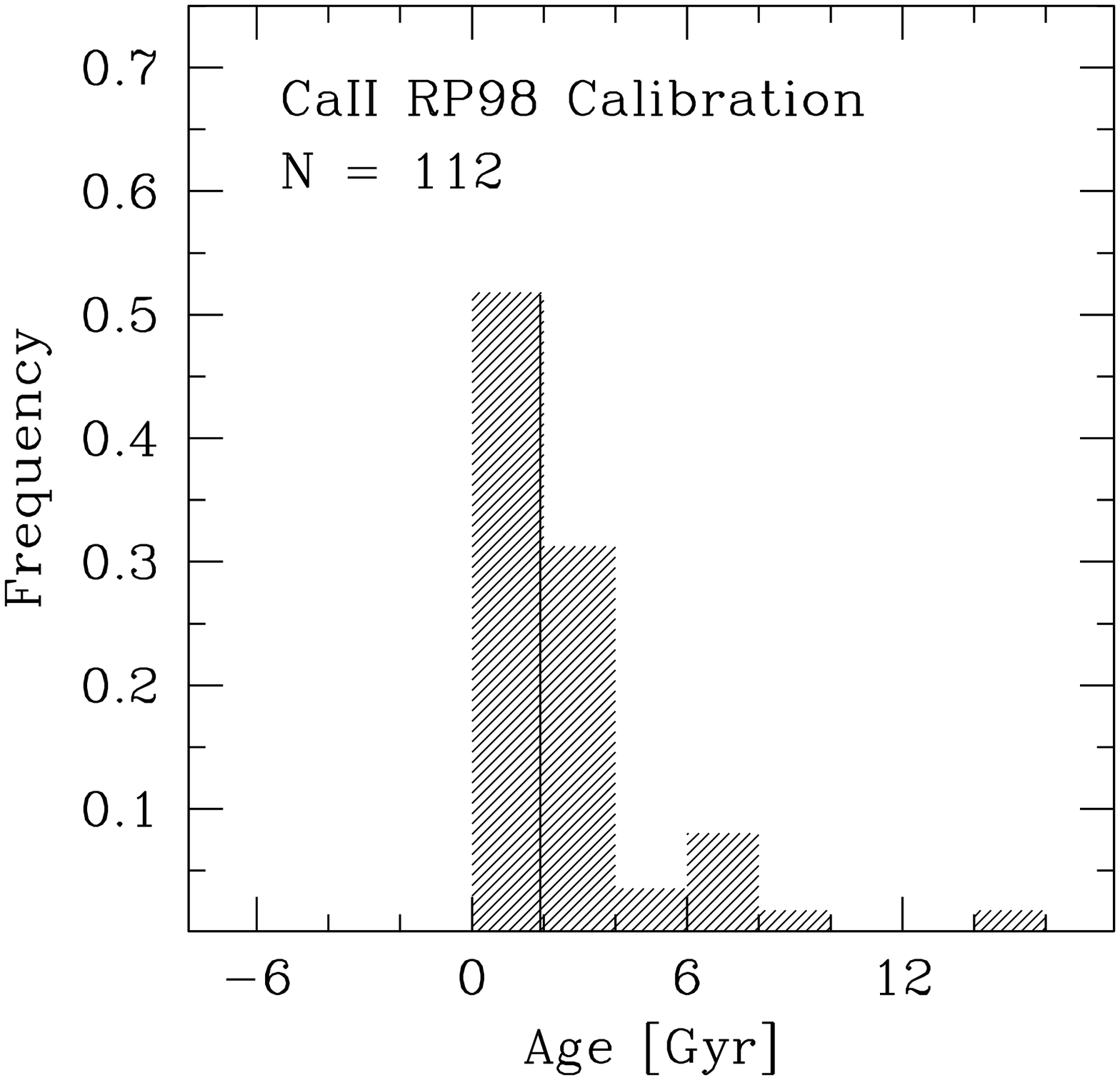}
\includegraphics[width=60mm]{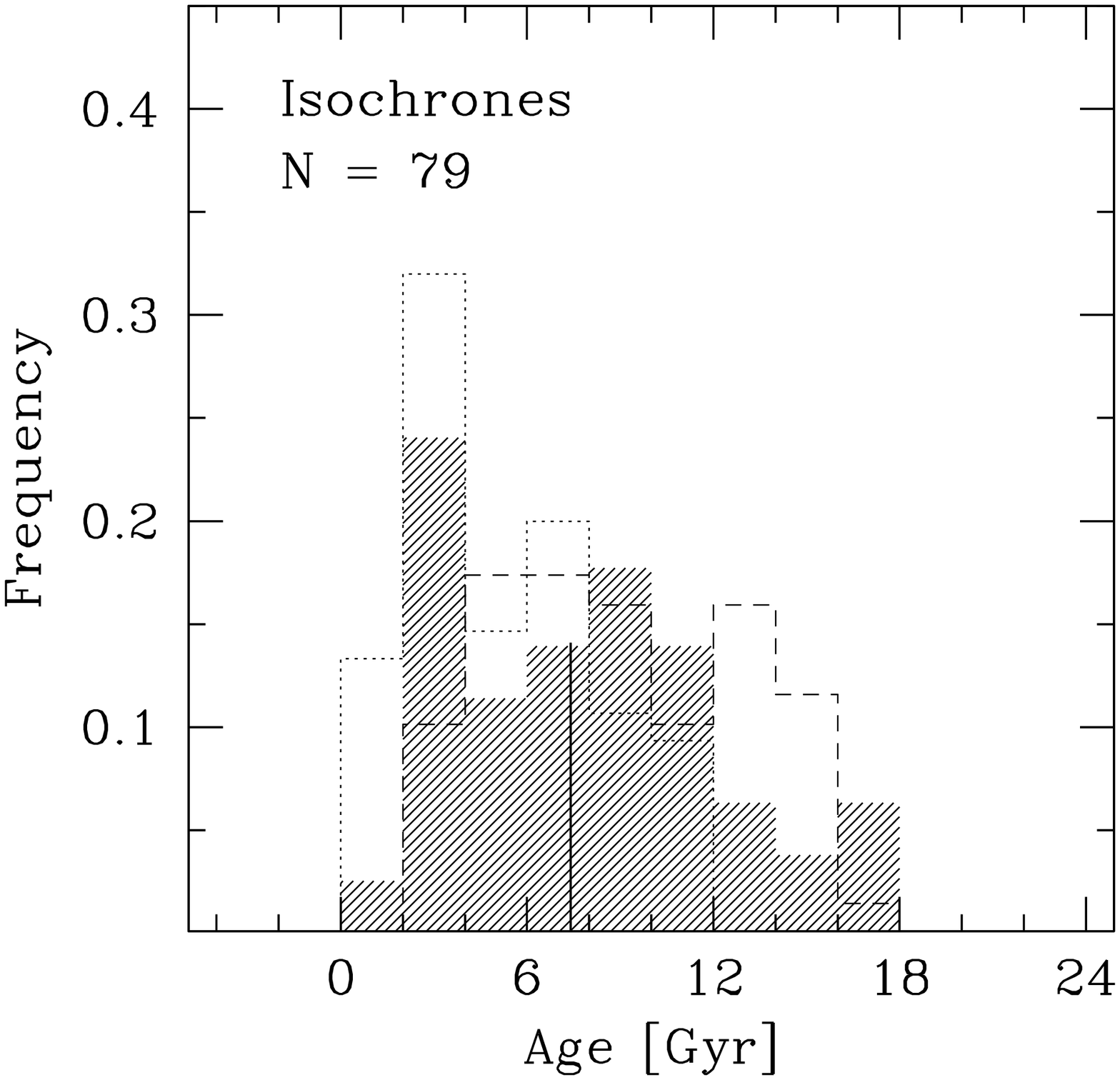}
\caption{Upper panels: Chromospheric age distributions derived from the
D93 and RPM98 calibrations. The lower panel correspond
to the isochrone distribution. Vertical continuous lines show the medians of each histogram.
In the isochrone age distribution we superpose with continuous
and dotted lines upper and lower limits estimations from \cite{nord}.}
\label{his1}
\end{figure}

Figure \ref{his1}, upper panels, shows the histogram distributions for
the ages derived using the chromospheric method applying D93 and RPM98
calibrations, respectively. In this section we analyze chromospheric ages
listed in Tables \ref{cromoages} and \ref{cromoagesnotcasleo}, i.e., including
objects with ages exceeding the 2 or 5.6 Gyr limits suggested by \citet{pace}
and \citet{wright1}, respectively (see section 3.).
In the isochrone age distribution (lower panel), we include upper and
lower limits with continuous and dotted lines, respectively, as derived by
\cite{nord}.  Li and [Fe/H] ages are not included in the histograms of Figure
\ref{his1}, as the methods used (see sections 4.2 and 4.3)
introduce bias toward younger and older ages, respectively.
Moreover the number of EH stars with Li age is relatively small (20 objects).

The ages distribution derived from the D93 calibration is broad.
On the contrary, the chromospheric age histogram obtained applying
the RPM98 calibration is quite narrow, with most of the objects having
ages $<$ 4 Gyr.  The isochrone age distribution is also quite broad with
two maxima at \hbox{3 and 9 Gyr}, respectively, probably showing the separation
between F and G spectral types within the EH sample.
\cite{kara} also noted that their distribution of
ages for the EH stars, derived using the isochrone technique, was flat or uniform
(having roughly the same number of stars in each age bin)
over 3 $\la$ age $\la$ 13 Gyr.  Table \ref{medianages} lists the medians and
the dispersions of the histograms in Figure \ref{his1}.

\begin{table}
\center
\caption{Medians and standard deviations for EH star age distributions}
\begin{tabular}{lccr}
\hline
 Method                     & Age [Gyr] & $\sigma$ [Gyr] & No. of stars\\
\hline
Chromospheric RPM98               & 1.9  & 4.0 & 112 \\
Chromospheric D93                 & 5.2  & 4.2 & 112 \\
\hline
Isochrone lower limit             & 4.8  & 3.0 & 75  \\
Isochrone                         & 7.4  & 4.2 & 79  \\
Isochrone upper limit             & 9.2  & 3.9 & 69  \\
\hline
\end{tabular}
\label{medianages}
\end{table}

The chromospheric age distributions resulting from the
application of the D93 and the RPM98 calibrations are very different,
with significantly different medians (see Table \ref{medianages}).
As mentioned before the RPM98 calibration includes a metallicity
correction factor whereas the D93 relation is independent of the
stellar metallicity effect.  For instance, for a EH star with
[Fe/H]$=$0.16, corresponding to the median of the EH sample \citep{santos},
the RPM98 calibration gives an age $\sim$ 3 Gyr lower (younger) than
the D93 relation. However, from \hbox{Figure \ref{his1}} and Table \ref{medianages}
it is clear that, the age distribution derived using the D93
calibration agrees better with the isochrone ages than the distribution
obtained from the RPM98 relation. 

\cite{felt} have suggested that the RPM98 metallicity correction
is not properly defined. The correction factor
for older ages is larger than for younger ages, as is given in
\hbox{$\Delta$Log age.}  We adopt the D93 calibration in spite of the relatively
high metal abundance of EH stars, in view of the poor determination of
this effect.

\begin{figure}
\center
\includegraphics[width=60mm]{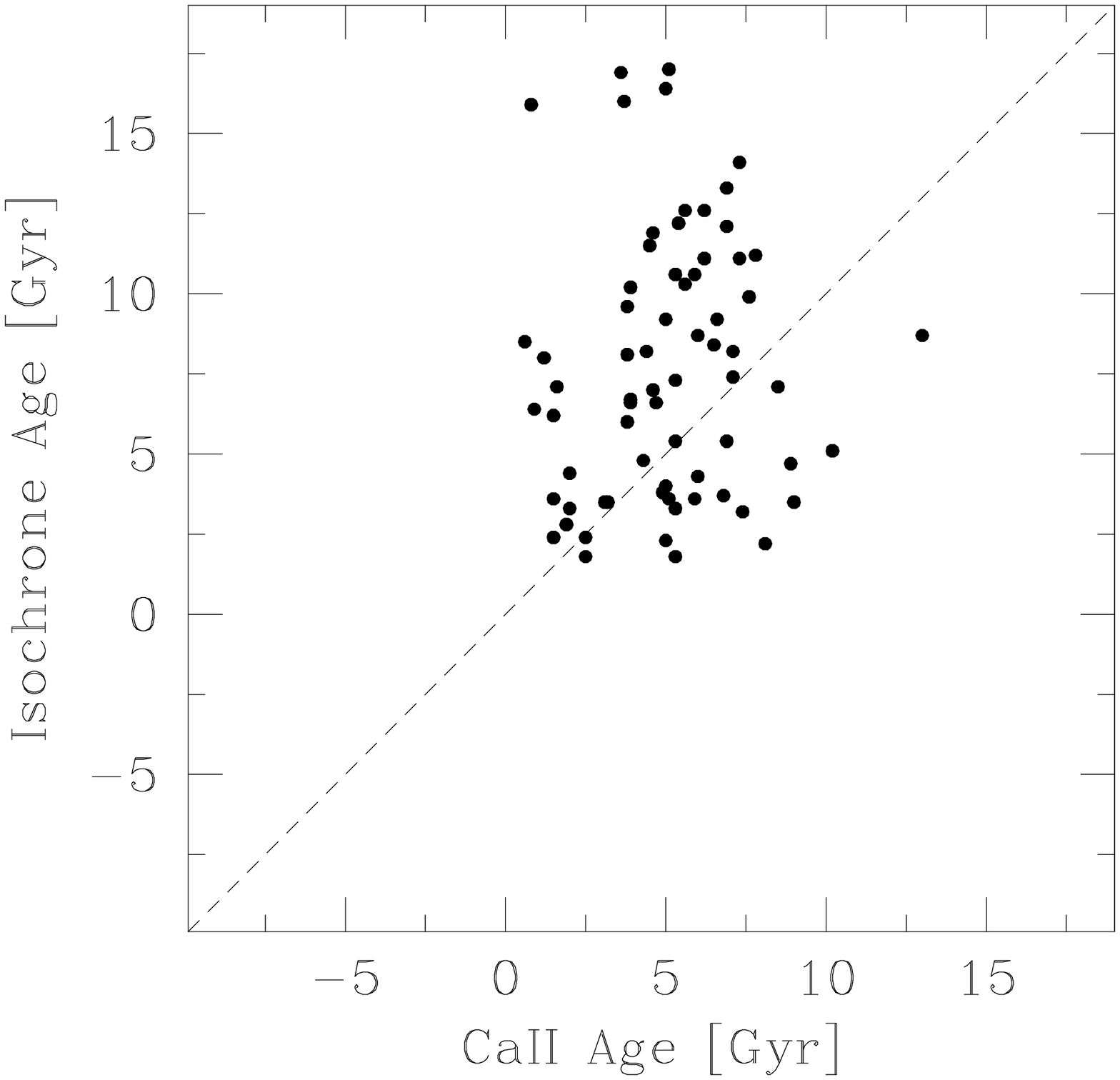}
\includegraphics[width=60mm]{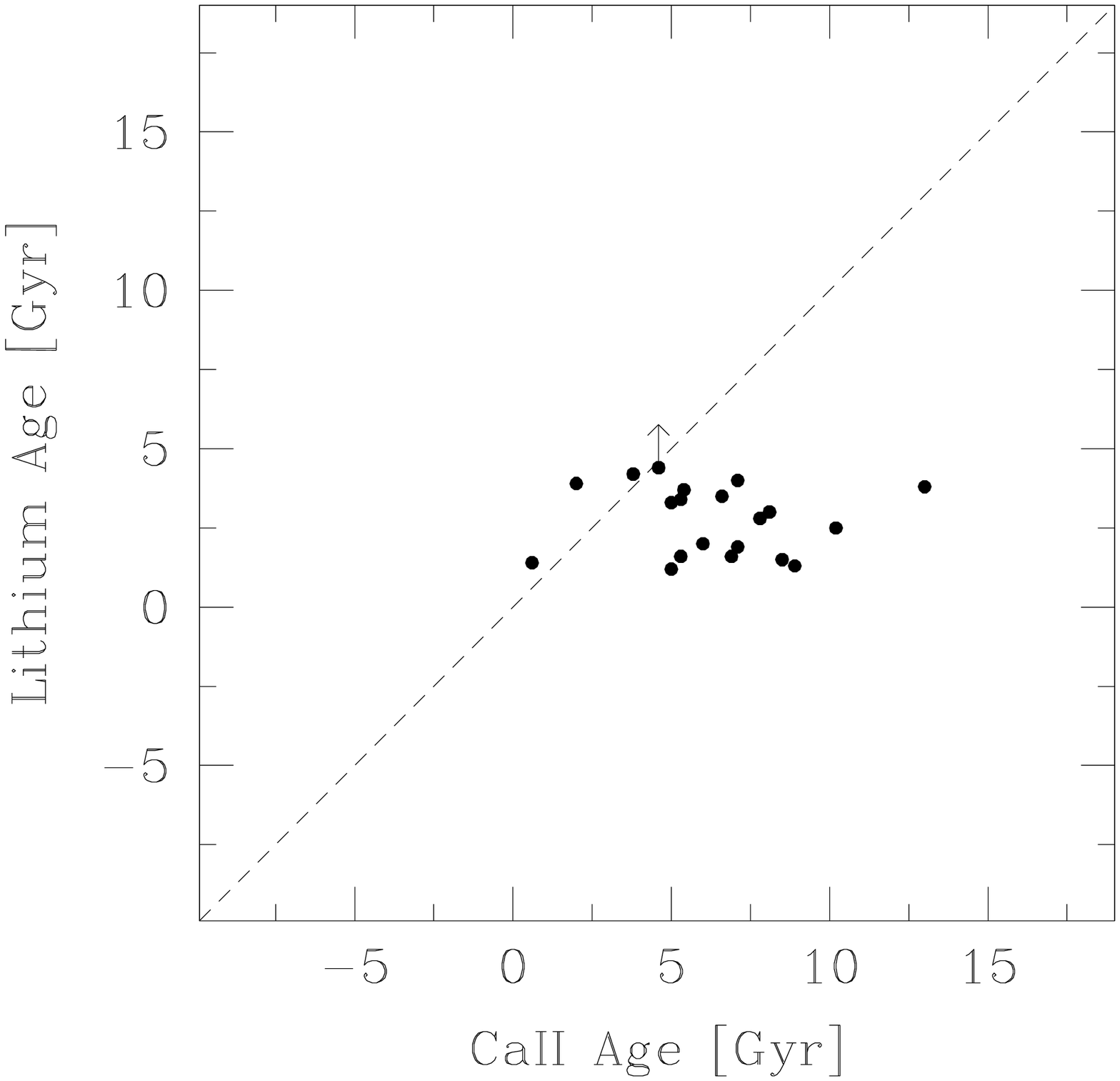}
\includegraphics[width=60mm]{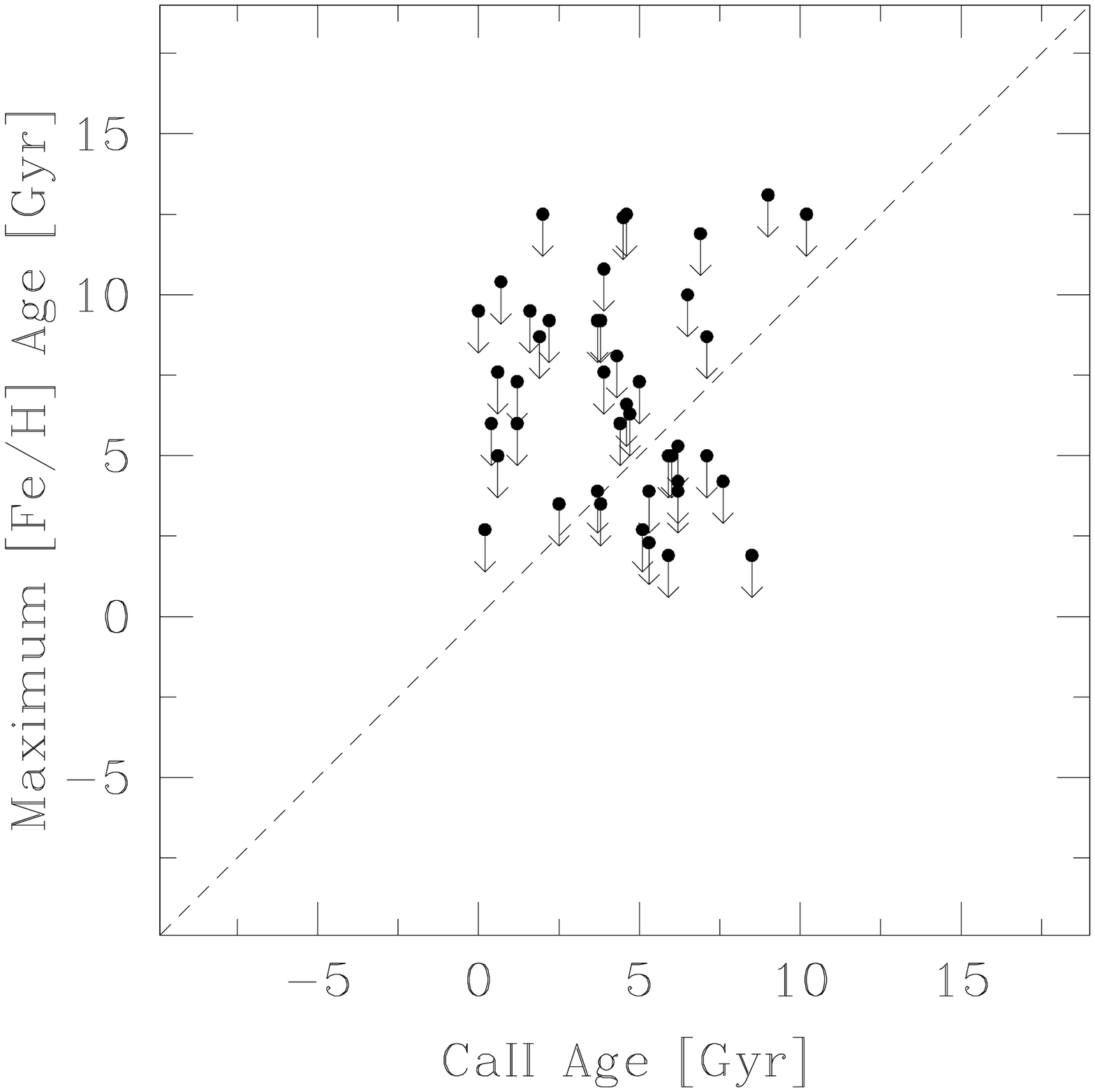}
\caption{Chromospheric ages against isochrone and lithium ages, and [Fe/H]
upper limits, respectively.}
\label{against}
\end{figure}

In Figure \ref{against} we plot the chromospheric ages vs the
isochrone and lithium ages. In addition we show the chromospheric ages
against the [Fe/H] upper limits.
Chromospheric ages are systematically smaller than isochrone
determinations and larger than lithium ages. However, the relatively
younger lithium ages are probably only reflecting the bias introduced by the
calibration used as discussed in section 4.2.
The metallicity derived ages are, on average, systematically older
than the chromospheric determinations (see Figure \ref{against}).
As for the Li ages, this is probably due to a selection effect
against metal-rich stars and thus younger objects, in this case
introduced by the metallicity-age relation (see section 4.3.).
We estimate a dispersion of about $\sim$ 4 Gyr for the chromospheric,
isochrone, metallicity (upper limit) ages.
The lithium age distribution in Figure \ref{against} (upper right panel)
has the smallest dispersion ($\sim$ 2 Gyr), however again this is probably due to the
fact that only the younger ages are taken in account. In conclusion, for the
Li ages, neither the the age difference (with respect to the chromospheric ages)
nor the relatively smaller dispersion are attributed to real features
but rather to the lack of older stars with Li ages estimations.
In the case of the [Fe/H] upper limits, the exclusion of younger objects restrain
a comparison with the chromospheric ages.

The chromospheric activity is a reliable age indicator for F and G
dwarfs from young ages to about $\sim$ 2.0 Gyr, adopting the most conservative
limit suggested by \cite{pace}, or possibly up to 5.6 Gyr, according to
\cite{wright1}'s result. On the other hand, isochrone ages are more precise
for stars that have evolved
significantly away from the ZAMS, up to 17 Gyr \citep[e.g.][]{nord}.
In this manner chromospheric and isochrone techniques are complementary
age estimator methods \citep{gus,lacha,felt,nord}.

The lithium method gives ages for only 20 of the EH stars, while the
ages derived from the metallicity technique are upper limits. The currently
available calibrations have bias against the older and younger stars, respectively.
In this sense they might also be considered complementary.
These two methods have greater uncertainties
than the chromospheric or isochrone methods \citep[e.g. ][]{gus,lacha,nord}
when applied to derive individual stellar ages.
Thus, the chromospheric and the isochrone techniques seem to be
the most reliable age indicators. In the case of the chromospheric
technique, the D93 calibration produces, in general, results in better agreement
with the isochrone technique than the RPM98 relation.

\subsection{Limitations of the different age estimators}

In this section we briefly comment on the applicability of the
different age estimators employed to derive ages for the EH stars,
and discuss systematic bias due to the use of these methods.

Chromospheric age determinations are based on
the chromospheric activity of F, G and K stars. It is
well established that the level emission decreases with
time. However, at the present time is not clear up to
what age this indicator can provide reliable stellar ages.
\citet{wright1} finds that the activity-age relation breaks
down for ages $\ga$ 5.6 Gyr while \citet{pace} determine
an earlier limit, of $\sim$ 2 Gyr. Nonetheless, this
relation has been used well beyond the 5.6 Gyr limit
\citep[see, for example][]{hen97,donahue,wright2}. In any event,
this method is more appropriate for the younger ages,
when the level of stellar activity is sufficiently high.
After this, the relation
breaks down and enters into a plateau, offering little or no
practical use as an age indicator.  The uncertainty in the ages
derived by this method depends on the temporal base line of
observation of each individual star
\citep[see, for example,][]{henry2000}. \citet{gus} estimated
a general uncertainty of about 30\% in the chromospheric age
derivations.

Isochrone ages strongly depend on the uncertainties of the
observables (T$_{\rm eff}$, M$_{\rm V}$, and metallicity).
In general the precision of
these ages strongly varies with the position of the star on the
HR diagram \citep[see, for example, ][]{nord}. For example,
for low mass objects the isochrone traces tend to converge
on this diagram \citep{pont}.
In addition, isochrone ages are less reliable for younger objects.
On the other hand, this technique provides relatively more precise ages
for stars that have evolved away from the ZAMS \citep{felt,lacha,nord}.
Recently, \citet{pont} have extensively discussed this issue
and proposed a method to estimate more accurate ages
based on the Bayesian probability. \cite{lacha} and \cite{nord}
have suggested typical uncertainties of about 50\% in the ages
derived by the isochrone technique.

Error estimations for isochrone and chromospheric ages are, in general,
relatively large, 30--50\%.  Figure \ref{against}
shows relatively large dispersions, particularly in the case of the
isochrone and chromospheric ages. This plot is probably reflecting
the uncertainties in both techniques and not a peculiarity of the EH sample
(see also Figure \ref{cross}).

The [Fe/H]-age relation has a large dispersion \citep{edv,carr}, although
\citet{pont} have significantly decreased this scattering. In addition,
currently available calibrations do not include metal-rich objects, introducing
a strong selection effect against (the younger) EH stars.
The Li-age calibration is poorly constrained \citep{soder2,pasq1,pasq2,boes91}
and biased toward younger objects.
Consequently the ages derived by these methods suffer from large
uncertainties. In spite of this, they are useful as independent
age estimators.

The kinematic technique can be applied to estimate the
ages for groups of stars and not to obtain individual stellar ages \citep{reid}.
In addition, as the EH sample lies to the right of Parenago's
discontinuity in the velocity dispersion (B$-$V) diagram
(see Figure \ref{kinemvega}), where old as well as young stars
can be found.  In any event, the kinematic ages of the EH
stars are convenient for comparison with other groups of objects.

\section{Comparison with the Solar Neighborhood stellar ages}

To compare the ages of the EH sample with those of stars in the
Solar Neighborhood with similar physical properties we selected three
groups of nearby objects. \citet{san2,newsan} provided a group of
\hbox{94 F--G--K} stars with no exoplanets detected by the Doppler technique;
31 of these stars have isochrone ages derived by \citet{nord}.
Sample A is composed of these 31 stars. Sample B contains 8684 F--G
objects with distances between 3 and 238 pc taken from \cite{nord}.
Sample C has 1003 F--G single stars within the same range of
distances as sample B, and snapshot (or instantaneous) values
of Log R$'$$_{\rm HK}$ derived by \cite{hen} and \cite{R33}.
Isochrone ages for samples A and B were obtained from \cite{nord}.
Chromospheric ages for sample C were derived by \cite{hen} and
\cite{R33} applying the D93 calibration. We include only chromospheric
and isochrone ages in our comparison as these are the most reliable
estimators (see Section 4.5). In addition, we apply both methods
for all stars independently of the ranges more appropriate for
each of them, as also mentioned in Section 4.5.,
to avoid systematic effects.

In particular the 2 Gyr limit of \citet{pace} for the applicability
of the chromospheric activity-age relation would
introduce a strong bias toward younger ages, similar to
the Li technique, with only $\sim$ 15\% of the EH sample
within this limit. In this case the only meaningful comparison with
the solar neighborhood would be of isochrone ages.

\begin{figure*}
\center
\includegraphics[width=60mm]{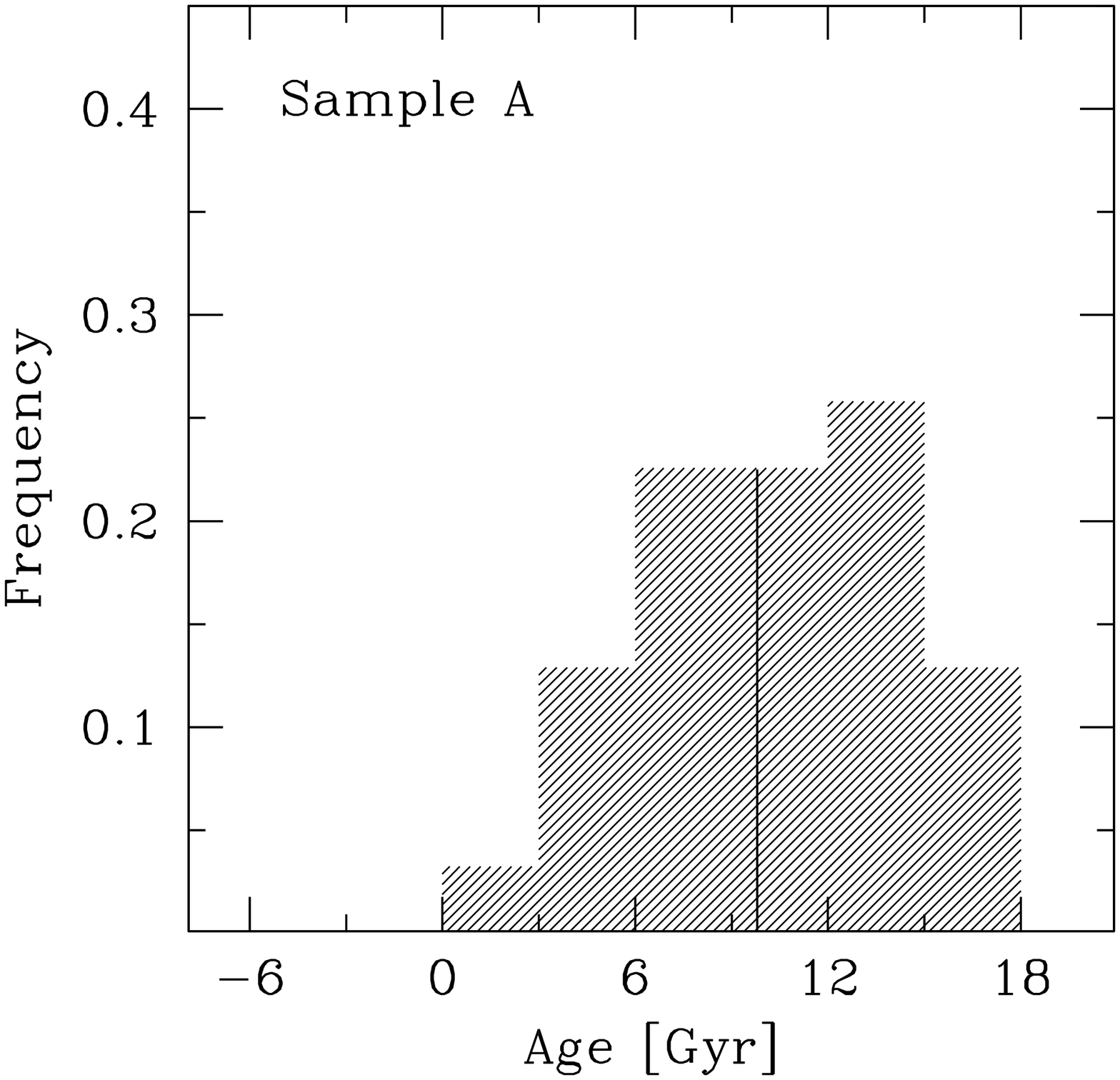}
\includegraphics[width=60mm]{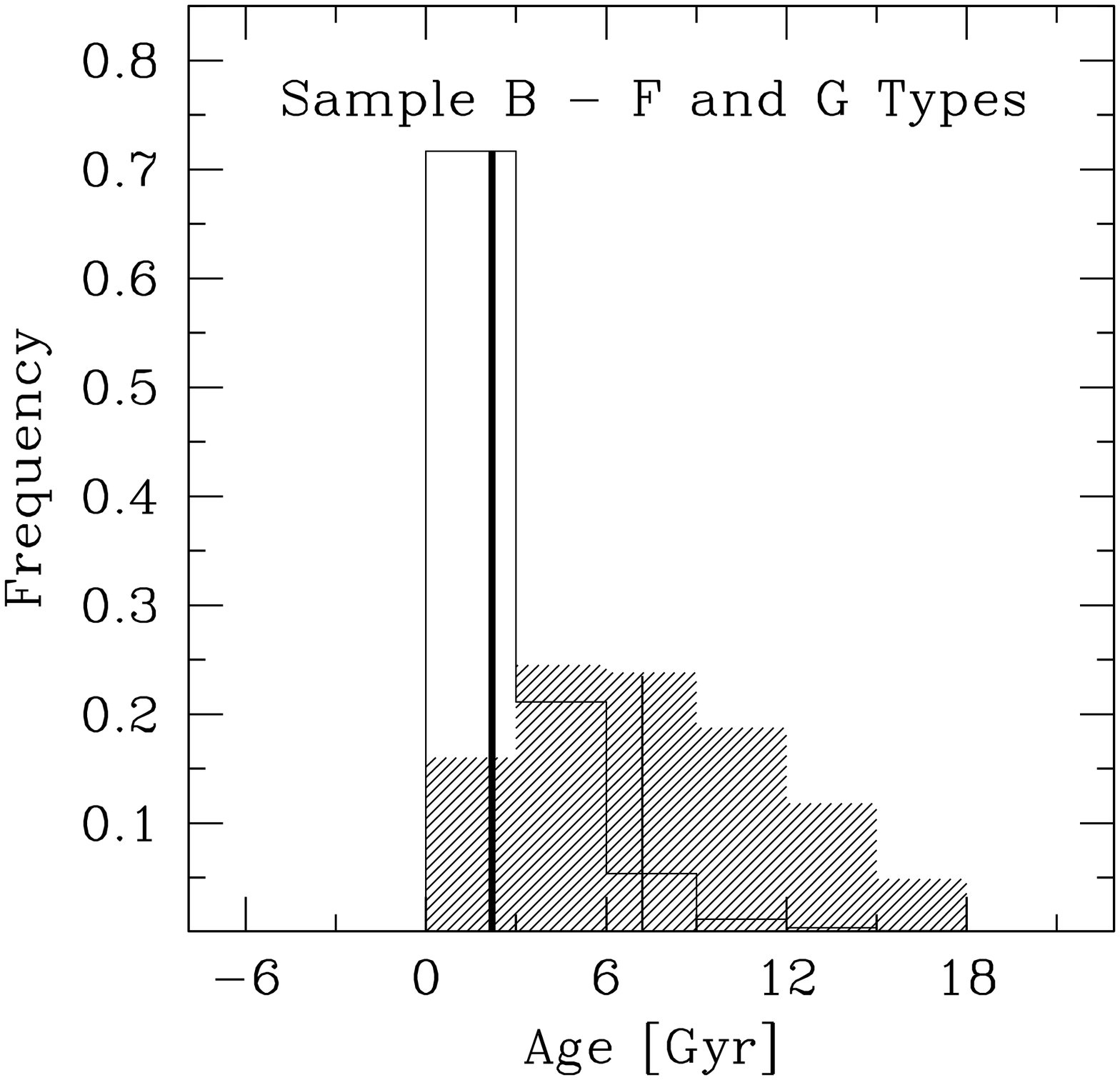}
\includegraphics[width=60mm]{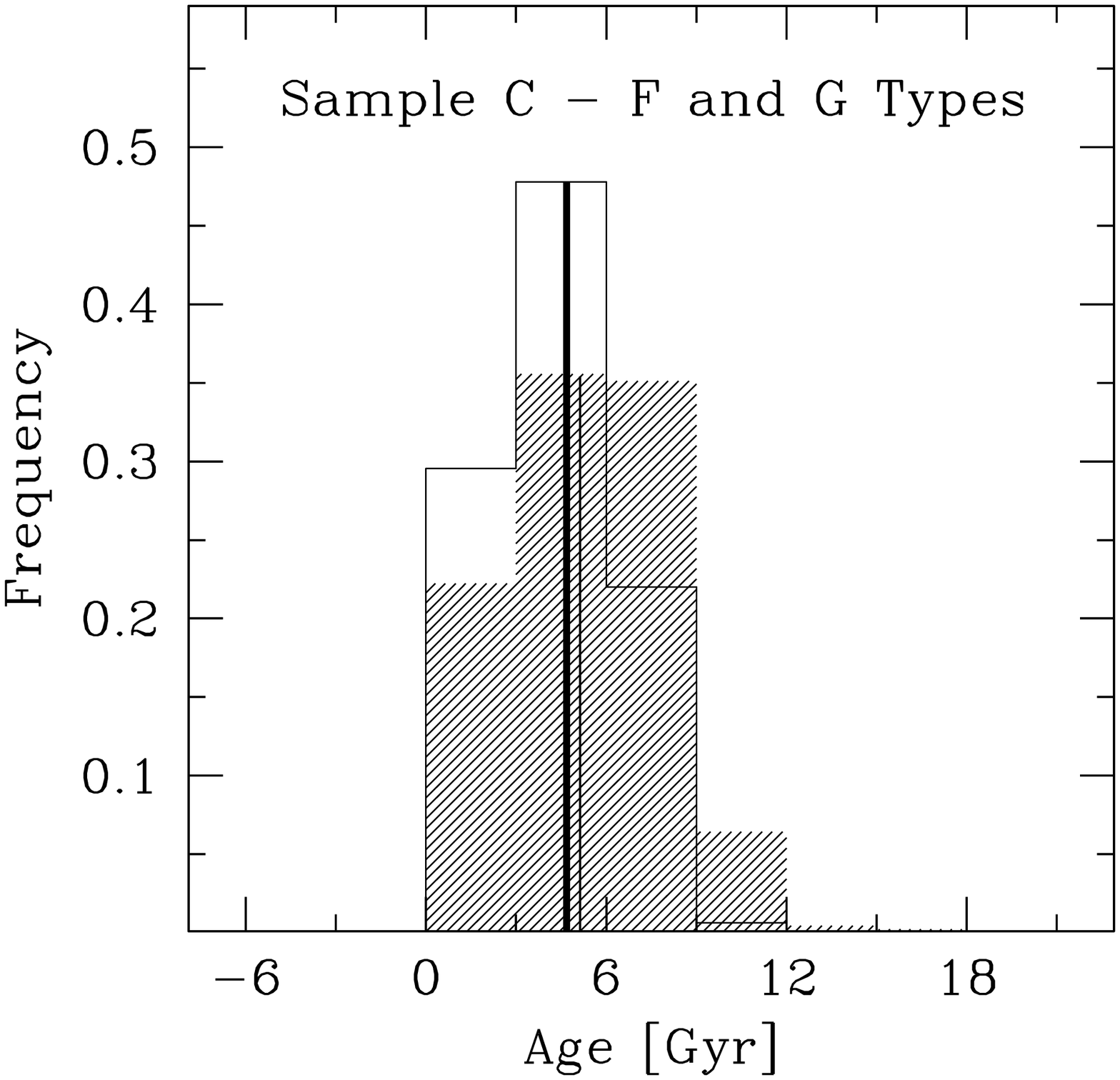}
\caption{Isochrone and chromospheric age distributions for
the Solar Neighborhood stars. Isochrone ages for
samples A and B were taken from \cite{nord}.
Chromospheric ages for sample C from
\cite{hen} and \cite{R33}. Spectral types G and F
in samples B and C are indicated by dashed and
empty histograms, respectively. The vertical continuous lines
show the median positions of each histogram. The thicker lines
correspond to the F type and the lighter to the G type.}
\label{agesfgk2}
\end{figure*}

\begin{figure}
\center
\includegraphics[width=60mm]{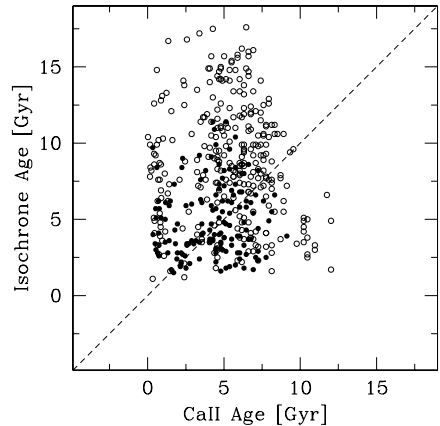}
\caption{Chromospheric ages vs isochrone ages for F and G-type solar neighborhood stars.
Filled and empty circles correspond to F and G spectral types, respectively.
Isochrone ages are taken from \cite{nord} and chromospheric determinations
from \cite{hen} and \cite{R33}.}
\label{cross}
\end{figure}

Figure \ref{agesfgk2} shows the isochrone age distributions for
samples A and B and the chromospheric age distribution for
sample C. In the cases of samples B and C the dashed and
empty histograms correspond to G and F spectral types, respectively.
We have cross-correlated samples B and C and plotted the result in
Figure \ref{cross}. A systematic effect is apparent in this figure.
Isochrone ages are, on average, larger (older) than chromospheric ages for the
Solar Neighborhood. This trend has already been seen in Figure \ref{against}
for the EH sample. We note that this systematic offset between
isochrone and chromospheric ages would also persist for
the 2 Gyr limit of \cite{pace}.

\begin{figure*}
\center
\includegraphics[width=60mm]{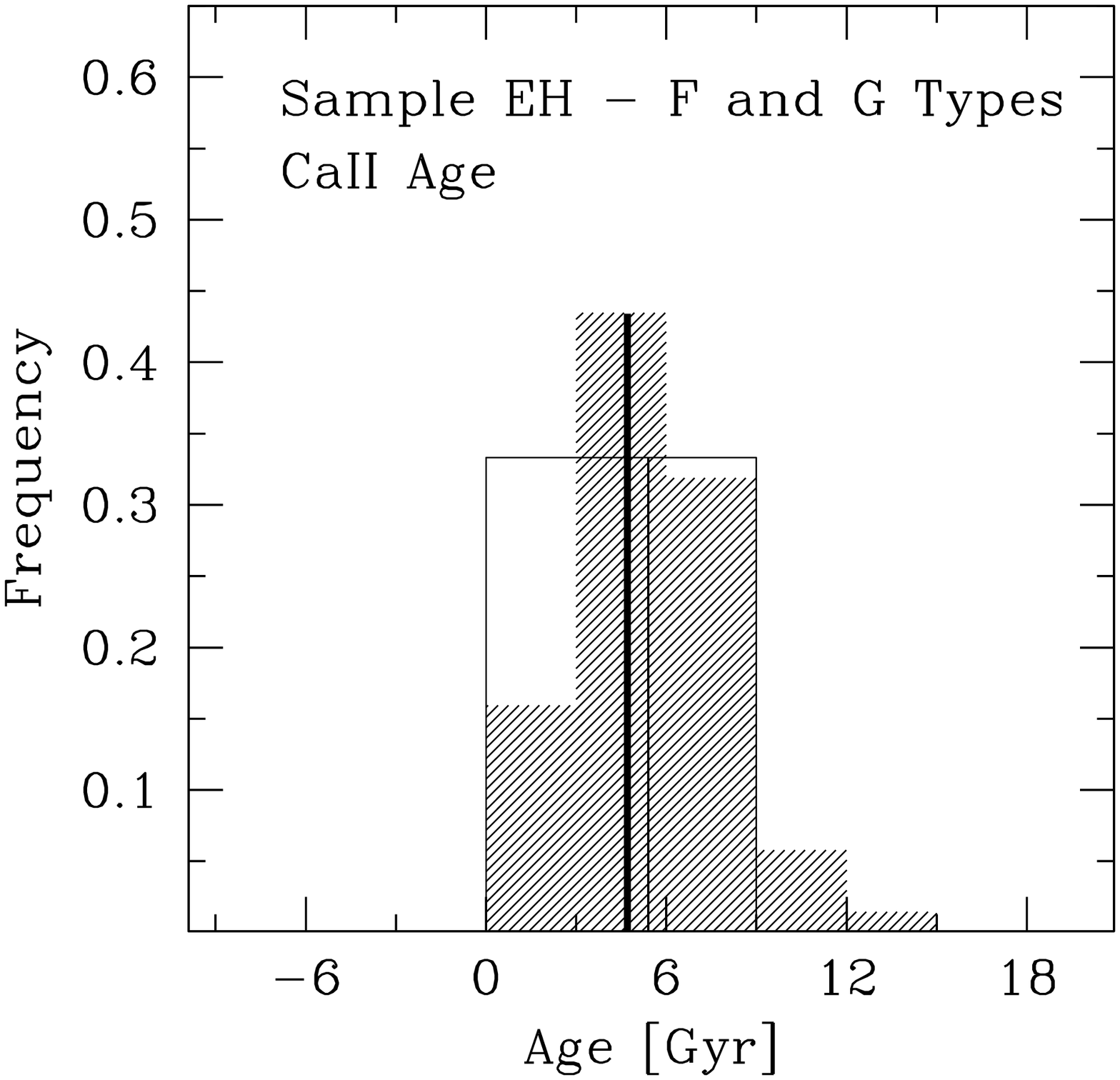}
\includegraphics[width=60mm]{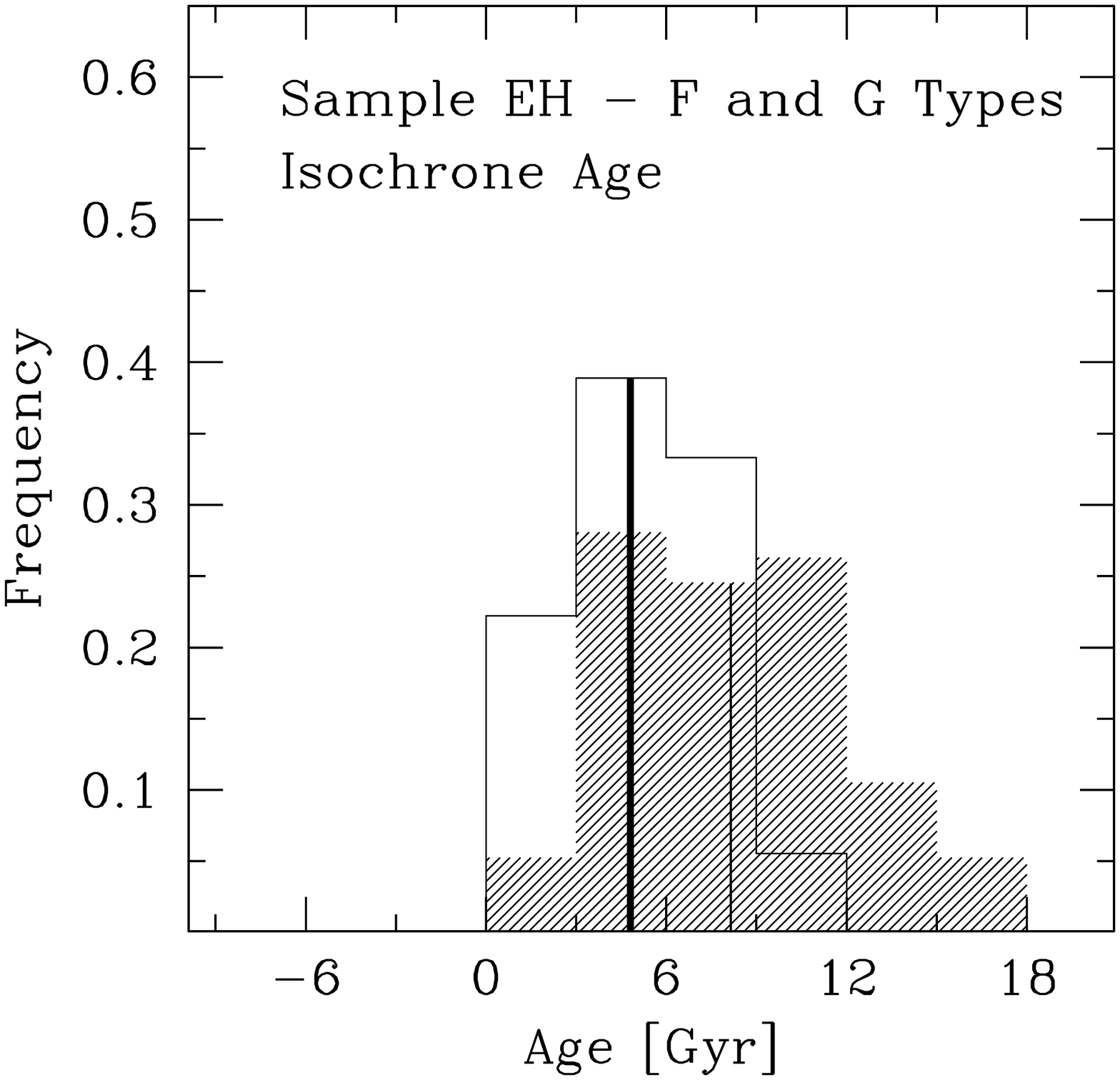}
\caption{Chromospheric and isochrone age distributions of EH stars.
Spectral types G and F are indicated by dashed and empty histograms, respectively.
The vertical continuous lines
show the median positions of each histogram. The thicker lines
correspond to the F type and the lighter to the G type. }
\label{agesfgk1}
\end{figure*}

In Figure \ref{agesfgk1} isochrone and chromospheric age distributions
for the EH group are shown by separating the sample in G and F stars.
Table \ref{tableagesfgk} gives the medians and the standard deviations
of samples A, B, and C as well as the EH group.
The isochrone technique gives different median ages for objects of
spectral types F and G for the nearby and the EH groups. On the other hand,
the chromospheric method does not discriminate by spectral type.
The median isochrone age for the (G and F) EH stars are $\sim$ 1--2 Gyr
larger (older) than for G and F stars in the solar neighborhood group. However
the dispersions are large, $\sim$ 2--4 Gyr.

The median age of G-type stars in sample A, 10.8 Gyr (25 objects, $\sigma$ $=$ 3.4 Gyr),
and the median age of the same spectral type stars in the EH group, 8.2 Gyr (isochrone ages),
are larger (older) than the median isochrone age for G-type stars in sample B, 7.2 Gyr
(see Table \ref{tableagesfgk}). This is probably reflecting the fact that stars
with high precision radial velocity measurements are selected among the less
chromospherically active and thus, on average, are relatively older objects
whereas sample B includes both active and inactive stars. We note, however, that
the age dispersions for the three groups are large, including young and old objects in
all cases.

\cite{beich} searched for infrared excesses in 26 FGK EH stars.  These
excess emissions are usually attributed to the presence of a debris disk
surrounding the central star. Whereas none of the objects show excess
at 24 $\mu$m, 6 of them do have excess at 70 $\mu$m. These authors,
among other analysis, compared the chromspheric ages
\cite[obtained from][]{wright2} for the
stars with planets with those of nearby stars not associated
with radial velocity detected planetary companions. They
derived median ages of 6 and 4 Gyr for the samples
of stars with and without exoplanets, although they considered
this difference not statistically significant. It is interesting
to note that we find a similar trend, with the EH stars being
1--2 Gyr older than the nearby stars. This result is based,
however, on the isochrone ages.

\begin{table*}
\center
\caption{Median ages for the EH and Solar Neighborhood stars}
\begin{tabular}{cccrccr}
\hline
Sample & Isochrone & Isochrone & Isochrone & Chromospheric & Chromospheric & Chromospheric\\
          & Median & $\sigma$ & N & Median & $\sigma$ & N\\
          & [Gyr]  & [Gyr]    &   & [Gyr]  & [Gyr]    \\
\hline
 Sample A          & 9.8   & 3.6  & 31   \\
\hline
 Sample B          & 2.8  & 3.7 & 8684 \\
F-type in Sample B & 2.2  & 2.0 & 5357 \\
G-type in Sample B & 7.2  & 4.1 & 2450 \\
\hline
 Sample C          & & & & 4.9  & 2.7 & 609 \\
F-type in Sample C & & & & 4.7  & 2.1 & 159 \\
G-type in Sample C & & & & 5.1  & 2.8 & 450 \\
\hline
EH sample           & 7.4  & 4.2 & 79      & 5.2  & 4.2  & 112 \\
F-type in EH sample & 4.8  & 2.6 & 18      & 4.7  & 2.4  & 15  \\
G-type in EH sample & 8.2  & 3.8 & 57      & 5.4  & 3.5  & 69  \\
\hline
\end{tabular}
\label{tableagesfgk}
\end{table*}

\begin{table*}
\center
\caption{KS statistical test results for the EH and the Solar Neighborhood stars}
\begin{tabular}{rrr}
\hline
                                & Isochrones ages  & Chromospheric ages \\
                                & KS test [\%]     & KS test [\%]      \\
\hline
EHs vs Sample A                & 0.92             &                  \\
\hline
EHs vs Sample B - F type only  & 5.3 $\times$ 10$^{-3}$    &          \\
EHs vs Sample B - G type only  & 50.4             &                   \\
\hline
EHs vs Sample C - F type only  &                  &  13.0              \\
EHs vs Sample C - G type only  &                  &  3.1 10$^{-5}$     \\
\hline
\end{tabular}
\label{ksneighbor}
\end{table*}

\begin{figure*}
\center
\includegraphics[width=60mm]{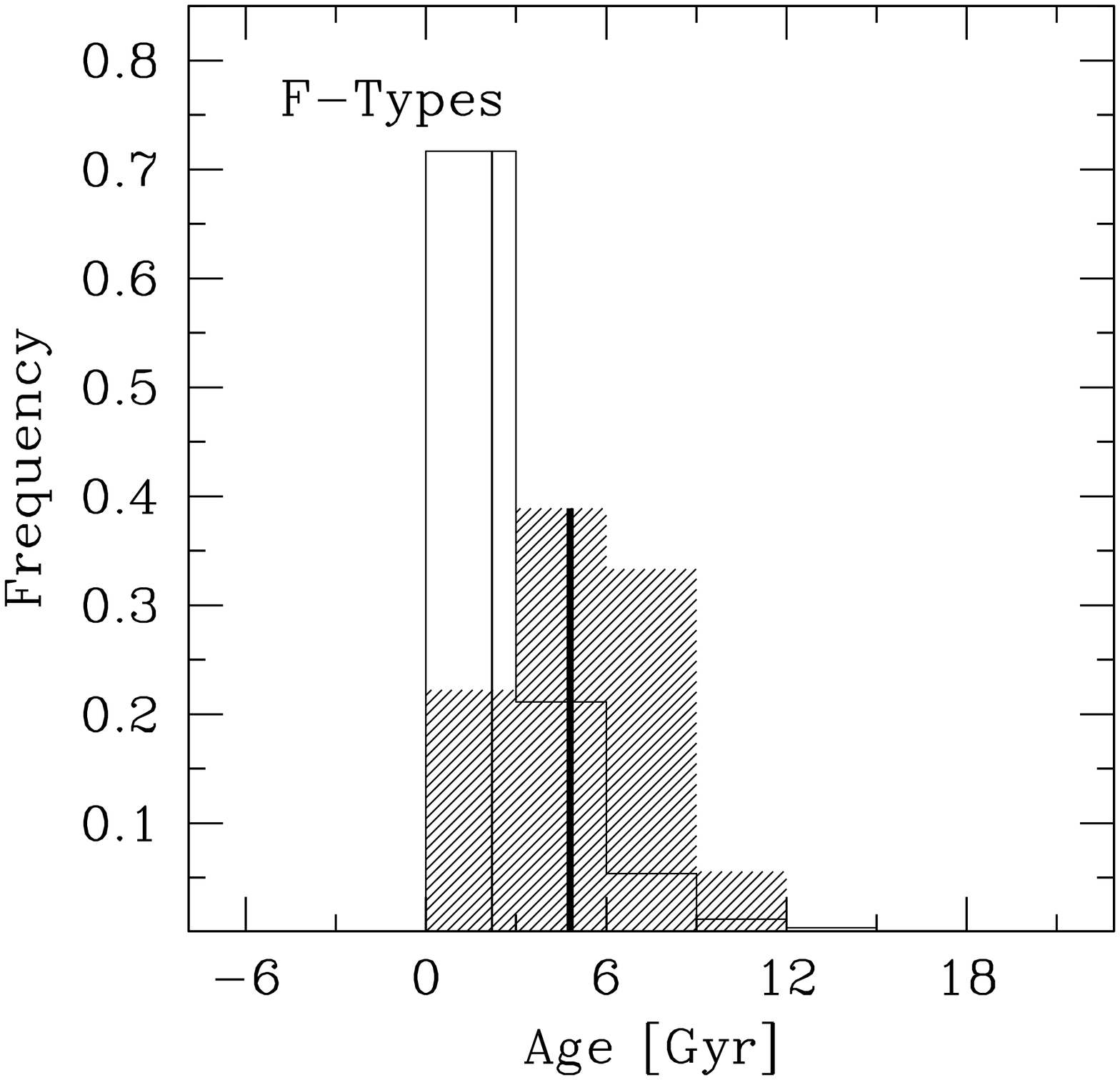}
\includegraphics[width=60mm]{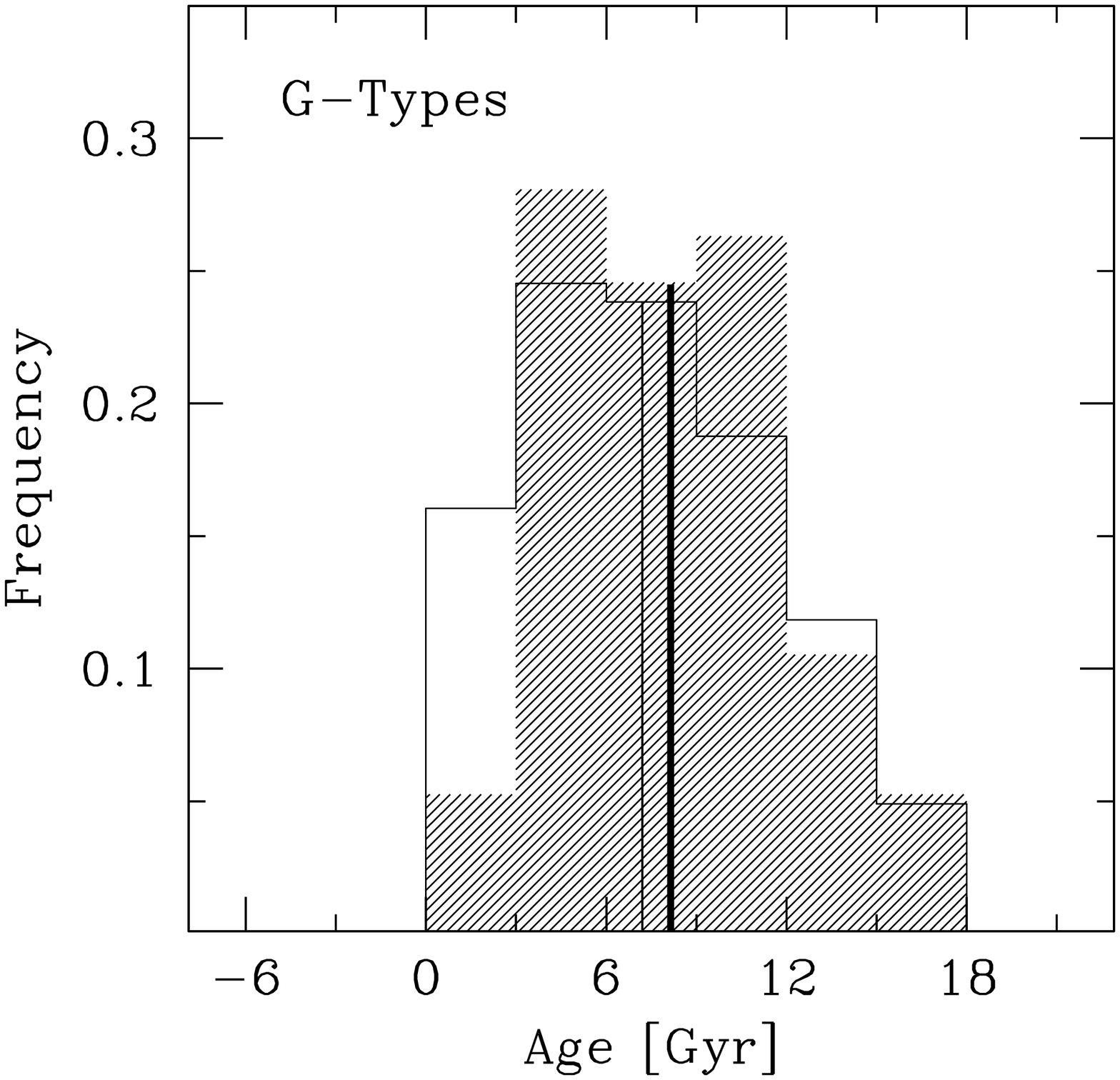}
\caption{Isochrone age distributions for EH stars (shade histograms) and sample B
(empty histograms). The right panel corresponds to stars of spectral type
F in both samples and the left panel to G-type objects.
The vertical continuous lines
show the median positions of each histogram. The thicker lines
correspond to the F type and the lighter to the G type.}
\label{super}
\end{figure*}

We applied the KS to compare the distributions in Figure \ref{agesfgk2}
with the EH histogram in \hbox{Figure \ref{agesfgk1}}.  Table \ref{ksneighbor}
shows the results.  The isochorone age distributions for
F type stars in sample B and in the EH group are different,
whereas for G type stars the distributions are more similar.
With respect to the chromospheric ages, stars of F and
G types in sample C and in the EH group show quite similar
distributions.

Figure \ref{super} shows the EH group and sample B
indicating stars of spectral types G and F in the left
and right panels, respectively.  The dashed histogram
corresponds to the EH distribution and the empty histogram
to sample B. In this figure, the apparent
age difference between EH stars and sample B stars, is
more evident for F spectral types. This is also
supported by the \hbox{KS test} result in Table \ref{ksneighbor}.
A similar comparison for the chromospheric
ages is meaningless as this method does not discriminate
by spectral types.

\section{Correlations of the stellar properties with age}

\begin{figure*}
\center
\includegraphics[width=60mm]{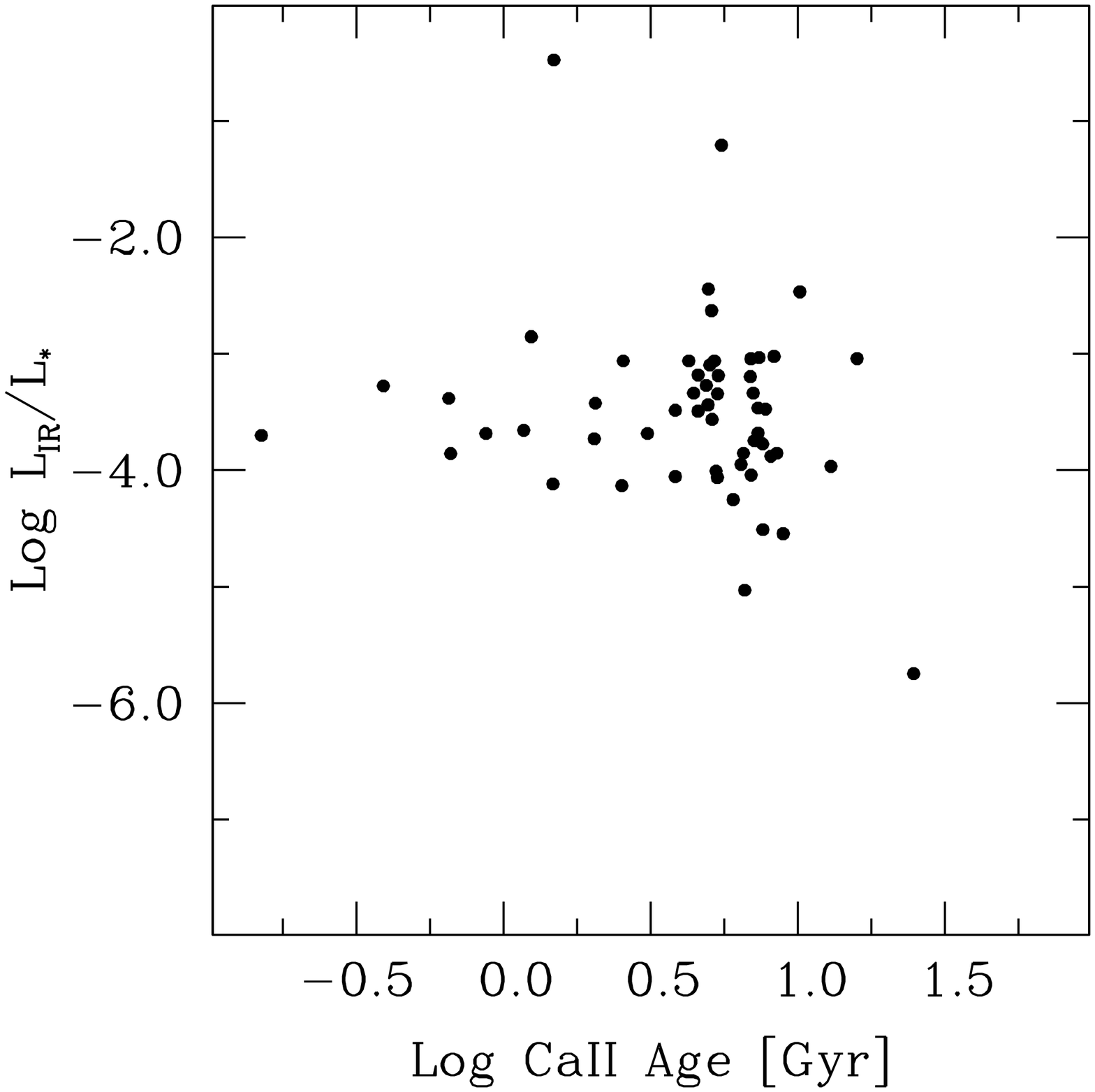}
\includegraphics[width=60mm]{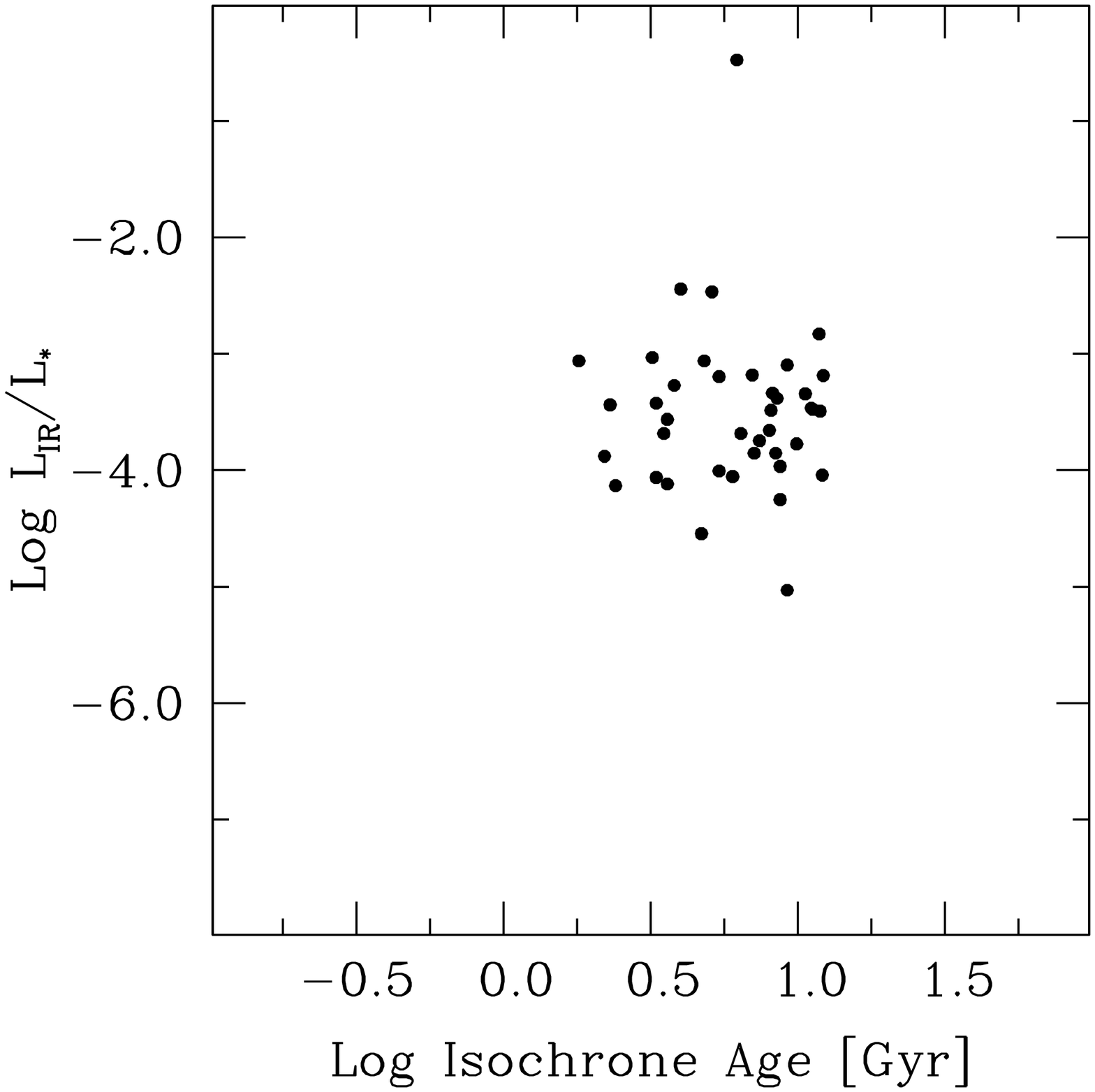}
\includegraphics[width=60mm]{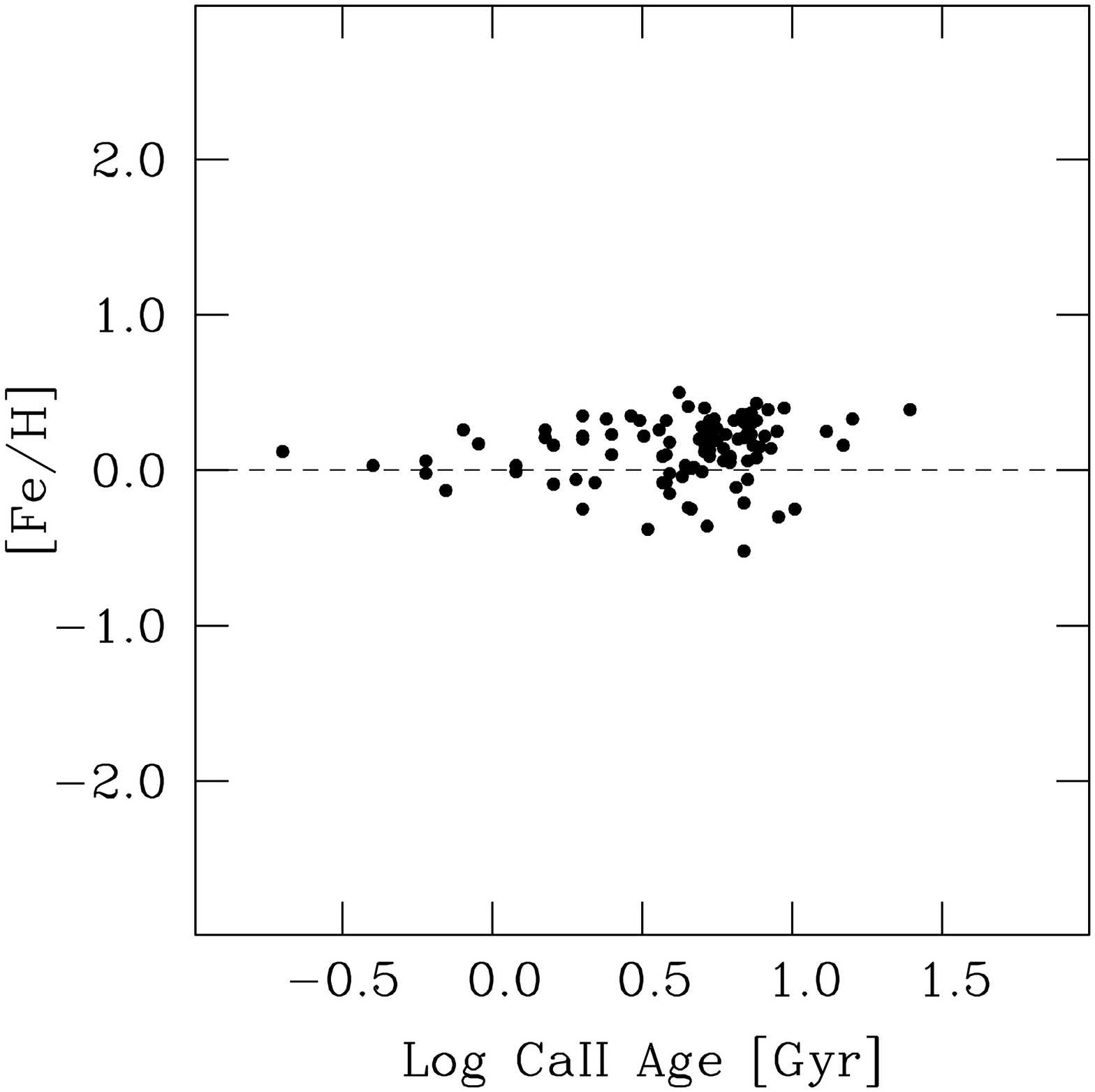}
\includegraphics[width=60mm]{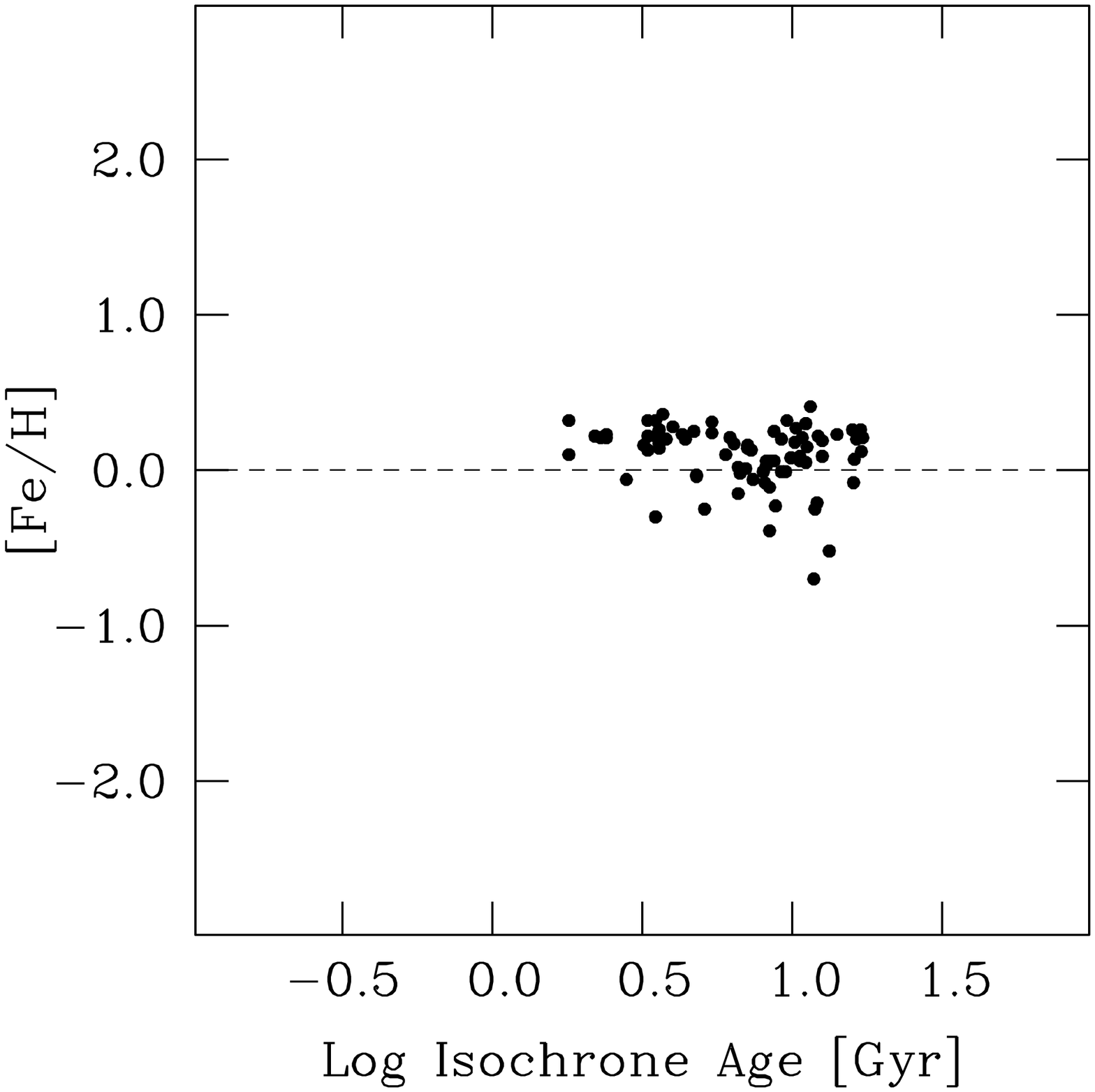}
\caption{Upper panels: L$_{\rm IR}$/L$_{\rm *}$ vs chromospheric and isochrone
ages. Lower panel: [Fe/H] vs chromospheric and isochrone ages.}
\label{correl1}
\end{figure*}

We searched for correlations between the EH stellar properties and age.
In Figure \ref{correl1}, upper panels, we plotted L$_{\rm IR}$/L$_{\rm *}$,
the excess over the stellar luminosity, vs chromospheric and isochrone ages.
The luminosity ratios
were obtained from \cite{saffe}. In the lower panels of the same figure
we show [Fe/H] vs age.
Whereas no correlation is apparent with the excess over the stellar luminosity,
a weak correlation shows up with the [Fe/H]. In other words, the metallicity
dispersion seems to increase with age.

To verify whether this is a real tendency,
we divided the EH sample in two bins, adopting \hbox{Log Age $=$ 0.5} as the
cut point for the CaII ages and \hbox{Log Age $=$ 0.75} for the isochrone
ages, to have more even sub-samples in both cases. We then calculated the
rms corresponding to the average age in each bin.
For the CaII ages (lower
left panel in Figure \ref{correl1}), we obtained
for Log Age greater and less than 0.5, an rms of 0.16 and 0.21 dex, respectively.
In the case of the isochrone
ages (lower right panel in Figure \ref{correl1}), we derived the same rms
for stars with Log Age in the first and second bin.
We repeated this analysis adopting chromospheric ages for
the EH stars with Log R$'$$_{\rm HK}$ $<$ $-$5.1 and isochrone
ages for the rest of the
objects. No substantial change is observed in Figure \ref{correl1}.

\cite{beich} found little or no correlation of the 70 $\mu$m
excess with the (chromospheric) age, metallicity and spectral type
of 6 EH stars. Nevertheless their analysis suggests that the frequency
of excess at 70 $\mu$m in stars with planets is at least as large
as for typical Vega-like candidate objects selected by IRAS \citep{plevy99}.
This result is somehow different to the finding of \cite{grea04} based on
sub-mm observations of 8 EH stars \citep{beich}.

\section{Summary}

We measured the chromospheric activity in a sample of
49 EH stars, observable from the southern hemisphere. Combining our
data with those from the literature we derived the chromospheric
activity index, R$'$${_{\rm HK}}$, and estimated ages for the
complete EH stars sample with chromospheric data 112 objects),
adopting the D93 calibration.
We applied other methods to estimate ages, such as: Isochrone,
lithium and metallicity abundances and space velocity dispersions,
to compare with the chromospheric results.

The derived median ages for the EH group
are 5.2 and 7.4 Gyr, using chromospheric and isochrone methods, respectively,
However, the dispersions in both cases are rather large, about $\sim$ 4 Gyr.
In the derivation of the median chromospheric age we have applied
the chromospheric activity-age relation beyond the 2 and 5.6 Gyr
suggested by \cite{pace} and \cite{wright1}, respectively.  In particular
the first limit would indicate that the isochrone technique is,
in practice, the only tool currently available to derive ages for the
complete sample of the EH stars.

 Lithium ages and metallicity upper limits are only available for
a subset of the EH stars, as the corresponding calibrations
do not cover the complete range of Li and [Fe/H] abundances
of this type of objects. This fact precludes any statistical
application of these ages for the EH stars.
The kinematic technique does not provide individual stellar ages.
In addition, kinematic ages are
less reliable because most EH stars lie to the right of Parenago's
discontinuity.

The median ages for the G and F EH stars derived from the isochrone method
are $\sim$ 1--2 Gyr larger (older) than the median ages for G and F solar
neighborhood stars.  We caution, however, that the dispersions in both
distributions are large, $\sim$ 2--4 Gyr.   The EH stars, analyzed here,
have been selected by means of the Doppler technique that favors the detection
of planetary-mass companions around
the less chromospherically active and slower rotator
stars, where radial velocity measurements can reach
high precisions of $\la$ few m/s \citep[see, for example,][]{hen97,R34,R40}.
As the chromospheric activity and the
rotation decrease with age, on average, we may
expect the EH stars be older than stars with similar physical
properties not known to be associated with planets.
The later group of objects is likely to include
a significant fraction of the more chromospherically active
and thus younger stars for which high precision in radial
velocity measurements are difficult to achieve.

With regard to the F EH stars, our result may suggest that
these objects are older than F nearby stars not known to be associated with
planets in opposition to \cite{sush}'s result. However, the relatively large
dispersion in our ages and the rather poor number of stars analyzed
by \cite{sush} render this apparent discrepancy meaningless.  We searched
for correlations between the age, the L$_{\rm IR}$/L$_{\rm *}$ and the
metallicity. No clear tendency is found in the first case, whereas
the metallicity dispersion seems to slightly increase with age.

\begin{acknowledgements}
This research has made use of the SIMBAD database, operated at CDS,
Strasbourg, France. An anonymous referee provided helpful comments
and suggestions that improved both the content and the presentation
of this article.
\end{acknowledgements}

\end{document}